# Magnetization signature of topological surface states in a non-symmorphic superconductor

W. J. Kuang[1], G. Lopez-Polin[1], H. Lee[2], F. Guinea[1], G. Whitehead[3], I. Timokhin[1], A. I. Berdyugin[1], R. Krishna Kumar[1], O. Yazyev[2], N. Walet[1], A. Principi[1]*, A. K. Geim[1,4], I. V. Grigorieva[1,4]*

[1]Department of Physics and Astronomy, University of Manchester, Manchester, UK
[2]Institute of Physics, Ecole Polytechnique Fédérale de Lausanne (EPFL), CH-1015 Lausanne, Switzerland
[3]Department of Chemistry, University of Manchester, Manchester, UK
[4]National Graphene Institute, University of Manchester, Manchester, UK

**Superconductors with non-trivial band structure topology represent a class of materials with unconventional and potentially useful properties. Recent years have seen much success in creating artificial hybrid structures exhibiting main characteristics of two-dimensional (2D) topological superconductors. Yet, bulk materials known to combine inherent superconductivity with nontrivial topology remain scarce, largely because distinguishing their central characteristic – topological surface states – proved challenging due to a dominant contribution from the superconducting bulk. Reported here is a highly anomalous behaviour of surface superconductivity in topologically nontrivial 3D superconductor $In_2Bi$ where the surface states result from its nontrivial band structure, which itself is a consequence of the non-symmorphic crystal symmetry and strong spin-orbit coupling. In contrast to smoothly decreasing diamagnetic susceptibility above the bulk critical field $H_{c2}$, associated with surface superconductivity in conventional superconductors, we observe near-perfect, Meissner-like screening of low-frequency magnetic fields well above $H_{c2}$. The enhanced diamagnetism disappears at a new phase transition close to the critical field of surface superconductivity $H_{c3}$. Using theoretical modelling, we show that the anomalous screening is consistent with modification of surface superconductivity due to the presence of topological surface states. The demonstrated possibility to detect signatures of the surface states using macroscopic magnetization measurements provides an important new tool for discovery and identification of topological superconductors.**

The study of materials with nontrivial topology, including topological insulators, semimetals and superconductors, is at the centre of current condensed matter physics research.[1-14] The recently developed theory of topological quantum chemistry[12-14] revealed that as many as a quarter of materials found in nature could possess nontrivial topology. This contrasts with the present status of experiment where relatively few materials – especially, among metals and superconductors - have been found to display tell-tale signs of nontrivial topology, i.e., topology-protected surface states.[3-6,15] Such states are particularly difficult to identify in metallic systems because their macroscopic properties are dominated by the bulk and the surface states' contribution is often negligible. This partially explains why only surface-sensitive techniques, such as angle-resolved photoemission spectroscopy[3,4,15] and tunnelling spectroscopy,[5,6] have been successful so far in detection of topological surface states in (semi)metals and superconductors.

Here we use an intrinsic superconductor $In_2Bi$ to demonstrate that topological surface states have a distinct signature in surface magnetization and can be detected above the bulk critical field for superconductivity, $H_{c2}$. The existence of surface states is predicted as a result of this material's nontrivial band structure, which itself is a consequence of the non-symmorphic crystal symmetry (implied by the space group P6$_3$/mmc) and strong spin-orbit coupling due to large atomic numbers of the constituent elements. The crystal structure of In$_2$Bi is illustrated in Figure 1a and can be viewed as a combination of two interpenetrating crystal lattices: a layered arrangement of In-Bi planes forming a hexagonal lattice in each layer (below we refer to these as In$_1$Bi$_1$) and a triangular array of 1D chains of In atoms piercing the centres of In$_1$Bi$_1$ hexagons. The screw symmetry of In$_2$Bi is associated with the AA' stacked monolayers of In$_1$Bi$_1$ whereas the In chains give the crystal its 3D character and presumably ensure little anisotropy in this material's properties.[16-18] In fact, the crystal symmetry of In$_2$Bi is similar to that of several known topological materials, including Dirac semimetal Na$_3$Bi,[15] non-symmorphic topological insulator KHgSb [19] and heavy-fermion odd-parity superconductor UPt$_3$ [20] where the surface states have been either predicted by theory or observed using surface-science techniques. Although basic superconducting characteristics of In$_2$Bi have been known for decades,[16-18] no attention had been paid to the nontrivial crystal symmetry and its consequences for the nature of superconductivity.

Importantly for the present study, we have succeeded in growing high-quality single crystals of In$_2$Bi, as confirmed by X-ray diffraction analysis (for details of the crystal growth and characterization, see Methods and Figures S1, S2 in Supporting Information). Both spherical and cylindrical samples of ~2 mm in diameter $d$ were studied, with all crystals exhibiting smooth, mirror-like surfaces (insets of Figure 1b and Figure S3). Below we focus on the results obtained for cylinders because of the simple geometry, best suitable for magnetization studies (spherical samples exhibited essentially the same behaviour described in Supporting Information). The high quality of our In$_2$Bi samples is also evident from the sharp (<0.1 K) superconducting transition at $T_c$ = 5.9 K, little hysteresis between zero-field cooling (ZFC) and field-cooling (FC) magnetization (Figure 1b), and nearly absent remnant magnetization $M$ (Figure 1c) indicating little flux trapping (pinning). The well-defined demagnetization factor for our crystals' geometry allowed us to accurately determine the characteristic parameters of In$_2$Bi superconductivity using the dc magnetization curves $M(H)$, such as shown in Figure 1c and Figure S3. At $T$ = 2 K (our lowest measurement temperature), we found the lower and upper critical fields $H_{c1} \approx 490$ Oe and $H_{c2} \approx 950$ Oe, respectively, coherence length $\xi \approx 60$ nm, magnetic field penetration depth $\lambda \approx 65$ nm and the Ginzburg-Landau parameter $\kappa = \lambda/\xi$ close to 1 (but see further for the temperature dependence of $\kappa$). In terms of these key superconducting parameters, In$_2$Bi is similar to very pure Nb,[21,22] one of the most-studied low-$\kappa$ superconductors, which allows comparison with conventional behaviour.

Figures 1c-d present our central observations: an anomalous magnetic response above the bulk critical field $H_{c2}$, where superconductivity is retained within a thin surface sheath of thickness ~2$\xi$ [23,24] and exists up to the critical field for surface superconductivity, $H_{c3}$. Surface superconductivity in conventional superconductors had been studied in much detail in the past, both theoretically and experimentally, and is known to have specific signatures in ac susceptibility $\chi = \chi' + i\chi''$ and dc magnetization $M(H)$ above $H_{c2}$. The contribution of the surface superconducting sheath to magnetization is particularly significant for pure superconductors with $\lambda \approx \xi$ [24-27], as in our case. As detailed in Supporting Information ('AC susceptibility and dc magnetization in conventional superconductors: Contribution of surface superconductivity'), above $H_{c2}$ the real part of ac susceptibility $\chi'$ is expected to evolve smoothly, decaying approximately linearly from the full Meissner screening at $H_{c2}$ to zero at $H_{c3}$. At the same time, $\chi''$ should exhibit a broad peak between $H_{c2}$ and $H_{c3}$ (Figure 1c). This standard behaviour has been well understood theoretically [24-29] as a consequence of shielding by the supercurrent induced in the surface sheath. The susceptibility is described by [28,29] $\chi' = \text{Re}\{[-1 + 2J_1(Kd)/KdJ_0(Kd)]/4\pi\}$ where $J_0$ and $J_1$ are Bessel functions and, ignoring the skin-effect and the contribution from normal electrons, $K^2 \approx -1/\lambda_L^2 = -4\pi n_s e^2/mc^2$. Here $\lambda_L$ is the London penetration depth, $n_s \propto |\Psi|^2$ the surface superfluid density, and $\Psi$ the order parameter.[30] The decrease in $\chi'$ with increasing applied field $H$ (blue curve in Fig. S10b) corresponds to a reduction of $n_s(H)$ inside the surface sheath (Fig. S10a) and a corresponding increase in $\lambda_L$, so that the screening ability of the sheath is gradually reduced. The broad maximum in $\chi''$ is due to normal electrons that appear above $H_{c2}$ and lead to dissipation, as they are accelerated by the electric field $E \propto dj_S/dt$ ($j_S$ is the supercurrent density). [31] Qualitatively, it can be explained as follows: as the dc field increases above $H_{c2}$, $n_S$ and $j_S$ are sufficiently large to cause an overall increase in dissipation as the density of normal electrons $n_n$ increases. However, as $n_S$ decreases further closer to $H_{c3}$, so does $j_S$ and $E$, which reduces the force on the normal fluid and the dissipation (at low frequencies used in our measurements the normal-state response is negligibly small). The expected $\chi'(H)$ and $\chi''(H)$ – which also reproduce the behaviour observed in pure conventional superconductors [21,22] – are shown in Figure 1c by the dashed blue lines (for further details, see Supporting Information and Figure S10).

In contrast to the described conventional behaviour, ac susceptibility of In$_2$Bi changes little above $H_{c2}$, showing near-perfect diamagnetism up to a certain, rather large, field $H_{ts}$ just below $H_{c3}$ (Figure 1c). There is a small decrease in $\chi'$ but otherwise In$_2$Bi exhibits a nearly complete Meissner effect with respect to the ac field. This is accompanied by vanishingly small dissipation $\chi''$ which indicates that the density of normal electrons remains negligibly small (see above). Only at $H_{ts}$, both $\chi'$ and $\chi''$ change abruptly, suggesting another phase transition, additional to the transitions at $H_{c1}$, $H_{c2}$, and $H_{c3}$. This anomalous behaviour becomes even clearer when we consider individual cycles of the magnetization, $m_{ac}(t)$, and the corresponding Lissajous loops $m_{ac}(h_{ac})$, where $h_{ac}(t)$ is the applied ac field (Figure 1d). Below $H_{ts}$ (yellow curves), $m_{ac}(h_{ac})$ are linear with 180° phase difference between $m_{ac}$ and $h_{ac}$, which indicates dissipation-free diamagnetic screening of the ac field. This behaviour persists up to $H_{ts}$ and is nearly identical to the full Meissner screening below $H_{c2}$ (blue curves). Only above $H_{ts}$, the ac susceptibility starts exhibiting the response normally expected for surface superconductivity: $m_{ac}$ decreases and out-of-phase signal appears so that the sinusoidal waveforms become strongly distorted (red and brown curves in Figure 1d) while $\chi'(H)$ smoothly decreases to zero at $H_{c3}$ and $\chi''(H)$ shows a corresponding maximum (Fig. 1c).

To observe this anomalous behaviour, it was essential to use very small ac fields. We could clearly see the transition at $H_{ts}$ in both $\chi'$ and $\chi''$ only using $h_{ac}$ below 0.1 Oe (Figure 2a). For larger $h_{ac}$ the additional features rapidly washed out, and only the standard behaviour could be seen for $h_{ac} \geq 1$ Oe (insets of Figure 2a). The phase transition at $H_{ts}$ was particularly clear for our smallest $h_{ac}= 0.01$ Oe (measurements became progressively noisier at smaller $h_{ac}$) where $\chi''$ split into two peaks and the shapes of $\chi'(H)$ and $\chi''(H)$ at $H_{ts}$ strongly resembled those observed near $H_{c2}$ but at much larger $h_{ac}$ (cf. curves for 0.01 and 1 Oe). This similarity serves as yet another indication of the new phase transition at $H_{ts}$. The observed sensitivity to the excitation amplitude is not surprising, as surface superconductivity is generally characterised by small $j_S$ and, therefore, can screen only small ac fields [27]. Furthermore, we found that the transition at $H_{ts}$ could be distinguished at all $T$ up to 5 K $\approx 0.85T_c$ (Figure 2b) and became smeared at higher $T$. The observed $T$ dependences for all three critical fields are shown in Figure 2d, where $H_{ts}(T)$ follows the same, almost linear, dependence as $H_{c3}$ (as expected,[32] the $H_{c3}/H_{c2}$ ratio is temperature dependent, with low-$T$ $H_{c3}/H_{c2} = 2.0$ decreasing to 1.69 at $T_c$, while $H_{c2}(T)$ for In$_2$Bi is linear in the available $T$ range; the linearity is discussed below).

The surface superconductivity of In$_2$Bi could be discerned even in our dc magnetization measurements (Figures 2c and S3b), which is unusual for a bulk superconductor, even for $\kappa \sim 1$: Firstly, this requires the presence of a continuous sheath of supercurrent which in bulk samples is typically interrupted by 'weak links' created by imperfections at the surface of realistic crystals; the weak links allow magnetic flux penetration and reduce the diamagnetic response.[27,33] Second, even in the ideal case, the corresponding dc signal at $H_{c2}$ is only $M_S \propto (H_c/\kappa)(\lambda/R)^{1/2}$ [25,27] ($H_c$ is the thermodynamic critical field and $R$ radius of the cylinder). In our case $4\pi M_S \approx 3G$ (Figure 2c), i.e., corresponds to the maximum theoretical value for In$_2$Bi parameters.[25] The contribution is diamagnetic if $H$ is increased, and paramagnetic for decreasing $H$, leading to a large hysteresis (Figure 2d). The hysteresis remained experimentally detectable in $H$ close to, but below $H_{ts}$. This behaviour is consistent with the presence of a continuous sheath of supercurrent at the surface, which prevents the magnetic flux from entering and exiting the normal-state bulk, leading, respectively, to a diamagnetic- and paramagnetic response.[25-27,33] The importance of the continuous sheath of current at the surface for the anomalous diamagnetism in our In$_2$Bi is further confirmed by its sensitivity to surface quality. When we intentionally introduced surface roughness, the anomalous features below $H_{ts}$ disappeared and the response became conventional with a smooth decrease of $\chi'$, a broad peak in $\chi''$ (Figure 3), and no hysteresis in $M(H)$ above $H_{c2}$, even though the critical fields $H_{c1}$, $H_{c2}$, and $H_{c3}$ were essentially unaffected. In contrast, bulk disorder was found to be less important for the anomalous behaviour: Bulk pinning reduced the diamagnetic susceptibility between $H_{c2}$ and $H_{ts}$ but the transition at $H_{ts}$ can still be seen in our In$_2$Bi crystals even with stronger pinning (Figure S4). This further emphasizes the importance of the surface for the observed anomalous screening.

To explain the highly anomalous diamagnetism between $H_{c2}$ and $H_{ts}$, let us first consider the electronic structure of In$_2$Bi. It is shown in Figure 4a as calculated using ab initio density functional theory and elucidated by tight-binding calculations (Supporting Information). Although the entire Fermi surface of In$_2$Bi is extremely complex with many sheets, one can immediately see one important feature of the electronic structure. The Fermi surface consists of cylinder-shaped parts extended along the $z$-axis, as well as rounded pieces. The former is a result of weakly coupled In$_1$Bi$_1$ planes that bring a 2D character whereas the rounded parts, indicating isotropic, 3D charge carriers, arise mostly from the In chains, as mentioned in the introduction. This combination of quasi-2D and 3D Fermi surfaces has profound implications for superconductivity and, in particular, explains the unusual linear $T$ dependence of $H_{c2}$ and $H_{c3}$ (Figure 2d). Such behaviour is in fact expected for multi-band superconductivity arising simultaneously from 2D- and 3D- type Fermi surfaces [34,35] (for details, see 'Temperature dependence of $H_{c2}$: fitting to the multiband theory' in Supporting Information). The multi-band superconductivity in In$_2$Bi and the importance of the contribution from 2D In$_1$Bi$_1$ sheets characterized by non-symmorphic symmetry are also corroborated by Figure 4b that shows pronounced changes in the shape of magnetization curves with increasing $T$. At low $T$, In$_2$Bi exhibits $M(H)$ typical for conventional type-II superconductors with low $\kappa$, but the dependence becomes borderline type-I closer to $T_c$ (see the curve at 5.6 K). This can be quantified [35] using the ratio of the Ginzburg-Landau (GL) parameter $\kappa_{GL}$ and the Maki parameter $\kappa_2$ obtained from the magnetization slope $dM/dH$ close to $H_{c2}$:

$$4\pi \frac{dM}{dH}\Big|_{H=H_{c2}} = \frac{1}{\beta_L(2\kappa_2^2 - 1)}$$

where $\beta_L = 1.16$. The Maki parameter $\kappa_2(T)$ found from our measurements is plotted in the inset of Figure 4b. For single-band superconductors, $\kappa_2(T)$ is known to vary little ($< 20\%$) with $T$ so that its value remains close to $\kappa_{GL}$. In our case, $\kappa_2/\kappa_{GL}$ changes by a factor of 2, which corresponds to a multi-band superconductor with Fermi surfaces having different symmetries,[34,35] in agreement with the electronic structure of In$_2$Bi.

Another essential feature found in our band structure calculations is Dirac-like crossings near H (and H') points in the Brillouin zone. This is shown in Figure 4c for the case of a finite width ribbon (full band diagram is provided in Figures S6, S8). The crossings are a result of the crystal symmetry of In$_2$Bi, which combines a screw-symmetry axis (C$_2$) and a 3-fold rotational symmetry axis (C$_3$) (Figure 1a). In particular, the C$_2$ screw symmetry effectively decouples the electronic states within individual In$_1$Bi$_1$ planes (for $k_z = \pi/c$, where $c$ is the inter-plane distance) and provides two copies of an "asymmetric" Kane-Mele model.[36]

Spin-orbit interaction (strong for In$_2$Bi) lifts the degeneracy of the corresponding Dirac-like bands at H (H') points, opening a large spin-orbit gap of about 0.5 eV, which – as is well known from literature [36-39] – hosts topological surface states for most surface terminations. Figure 4c shows representative results for a zig-zag termination, as described in the Supporting Information ('Topological surface states'). Our DFT calculations (Fig. S6) show that, in pristine In$_2$Bi, the Fermi energy crosses the Dirac-like bands near their touching point and, therefore, crosses the surface states as well. In the case of In$_2$Bi, these states are confined to a few atomic layers at the surface (Figure S9) and, as we show below, should have a profound effect on surface superconductivity, consistent with the experimental observations.

To evaluate the effect of the topologically protected ultrathin layer on the overall diamagnetic response, we note that the surface states are expected to couple with bulk superconductivity and also become superconducting by proximity,[40,41] creating an "outer-surface" superconducting layer. To account for this coupling, we have extended the standard Ginzburg-Landau description of surface superconductivity [24] to include the proximitized surface states that are modelled as a superconducting film of thickness $d \ll \xi, \lambda$. As detailed in Supporting Information ('Effect of topological surface states on surface superconductivity'), this film effectively 'pins' the amplitude of the order parameter at the surface to its maximum value $F = 1$ for all $H < H_{c3}$, see Figure S11a (this is to be compared with a gradual suppression of $F(0)$ by $H_{c2} < H < H_{c3}$ for standard surface superconductivity, Figure S11a and ref. [24]). Due to coupling between this $H$-insensitive outer sheath and the standard surface superconductivity, Cooper pairs in the overall ~$2\xi$ thick surface layer become much more robust with respect to pair-breaking by the magnetic field, and the superfluid density $n_s$ remains at ~70% of its maximum value even at $H \approx H_{c3}$ (Figure S11b). Figure 4d shows the calculated evolution of $\chi'(H) \propto n_s$ between $H_{c2}$ and $H_{ts}$, which is different from the conventional response but in agreement with the experiment. The model also allows us to understand other features of the anomalous response below $H_{ts}$. First, its exceptional sensitivity to $h_{ac}$ (Figure 2a) can be related to a finite depairing current density $j_0$ within the outer-surface layer. Indeed, $j_C$ is given by the thermodynamic critical field (or the superconducting gap) [30] and typically is ~$10^{10}$-$10^{11}$ A m$^{-2}$. Because most of the screening current flows within the 1-nm thick outer-layer (where $F$ is maximised, Figure S11a), it is straightforward to estimate that the layer can sustain only $h_{ac} \lesssim 1$ Oe, in good agreement with experiment. Note that the standard surface superconductivity can support similar $j_C$ but the current flows through a much thicker (~$2\xi$) layer and, therefore, should sustain proportionally larger $h_{ac}$. Second, $\chi''$ depends on the number of normal electrons contributing to dissipation, as discussed above. For nearly constant $n_s$ between $H_{c2}$ and $H_{ts}$, the corresponding $\chi''$ should be negligibly small compared to conventional surface superconductivity, which explains little dissipation below $H_{ts}$ (Figures 1c,2a). Finally, the transition at $H_{ts}$ probably corresponds to a switch of the outer-surface layer into the normal state. This is largely expected as the outer-surface superconductivity is proximity-induced and, therefore, should have a smaller gap than the intrinsic one and be destroyed at some field $H = H_{ts} < H_{c3}$. Above this field, only the normal surface superconductivity provides diamagnetic screening.

We note that the above model does not invoke the topological nature of the surface states and in principle could be realised if a 'conventional' atomically thin metallic layer were present at the surface. Such a trivial scenario, however, discounts two important facts: Firstly, the strong diamagnetic response is observed in *all* our samples with a good degree of crystal purity, either cylindrical or spherical. This suggests that the surface metallic state responsible for modifying the behaviour of |Ψ| at the surface is a robust feature of In$_2$Bi and cannot be the result of e.g. trapping by some random surface potential or another artefact. Secondly, the Meissner-like screening of the whole volume of our In$_2$Bi crystals above $H_{c2}$ requires a *continuous* sheath of supercurrent at the surface. This is a stringent condition that usually cannot be met in bulk superconductors due to inevitable surface imperfections [33]. In topological materials surface defects – as long as they are non-magnetic – do not cause backscattering and do not disrupt topologically protected counter-propagating surface currents because the existence of the topological surface states depends only on the global symmetry properties of the crystal, not on local properties of the surface. This follows from the general concept of topological states and was confirmed in experiments on topological insulators, e.g., in ref. [43], where the topologically protected surface-conducting sheath was shown to envelop the entire surface of a crystal, despite rough surfaces with stacked edges, steps and different terminations. The presence of symmetry-protected topological surface states in In$_2$Bi offers a natural explanation for the high degree of reproducibility and robustness of the continuous sheath of supercurrent. A further indication of the topological nature of the superconducting outer-sheath in our experiments is given by the complete suppression of the anomalous diamagnetism by surface roughness (Figure 3). This is likely to be caused by introduced point-like defects (e.g., vacancies), which have been shown [42] to result in localised states and interaction-induced magnetic moments, similar to the effect of point defects in graphene. The latter introduce backscattering of the topological states and should strongly suppress their (super)conductivity.[42]

Finally, attributing the enhanced diamagnetism to an accidental metallic sheath at the surface contradicts our other observations, such as the effect of allowing a thin layer of In$_5$Bi$_3$ ($T_c \approx 4.2$K) to form at the surface of In$_2$Bi (see 'Evidence of In$_2$Bi oxidation in air and the importance of surface protection' in Supporting information). In the temperature interval between 4.2 and 5.9K this corresponded to enveloping superconducting In$_2$Bi in a thin (submicron) layer of normal metal. In stark contrast to our main observations in Figures 1 and 2, this resulted in an almost complete suppression of $\chi'$ at $T > 4$K and a sharp reduction of $H_{c3}$, i.e., an effect opposite to the described contribution of the superconducting topological states. Neither can the observed strong diamagnetism below $H_{ts}$ be explained by the standard surface superconductivity that is somehow non-uniform, e.g., due to a slightly varying stoichiometry of the crystals at the surface. Firstly, we carefully checked the structure and chemical composition of our crystals before and after the measurements, and these remained unchanged. More importantly, non-uniformity always leads to an increased pinning which would reduce the surface diamagnetic screening rather than enhancing it (Figure S4), again in contrast to our observations.

The above model based on proximity-induced superconductivity of topological surface states qualitatively explains all the main features seen experimentally. Nevertheless, a more quantitative understanding is certainly desirable, which should take into account the unconventional symmetry of the topological states' pairing wavefunction and consider self-consistently their coupling to bulk superconductivity, beyond the phenomenological Ginzburg-Landau theory. Independently of the microscopic mechanism, the observed enhanced surface diamagnetism can be employed to probe possible topological superconductors and, if found, our results show that effects of topological superconductivity can be isolated from the obscuring conventional contribution from the bulk by using magnetic fields above $H_{c2}$.

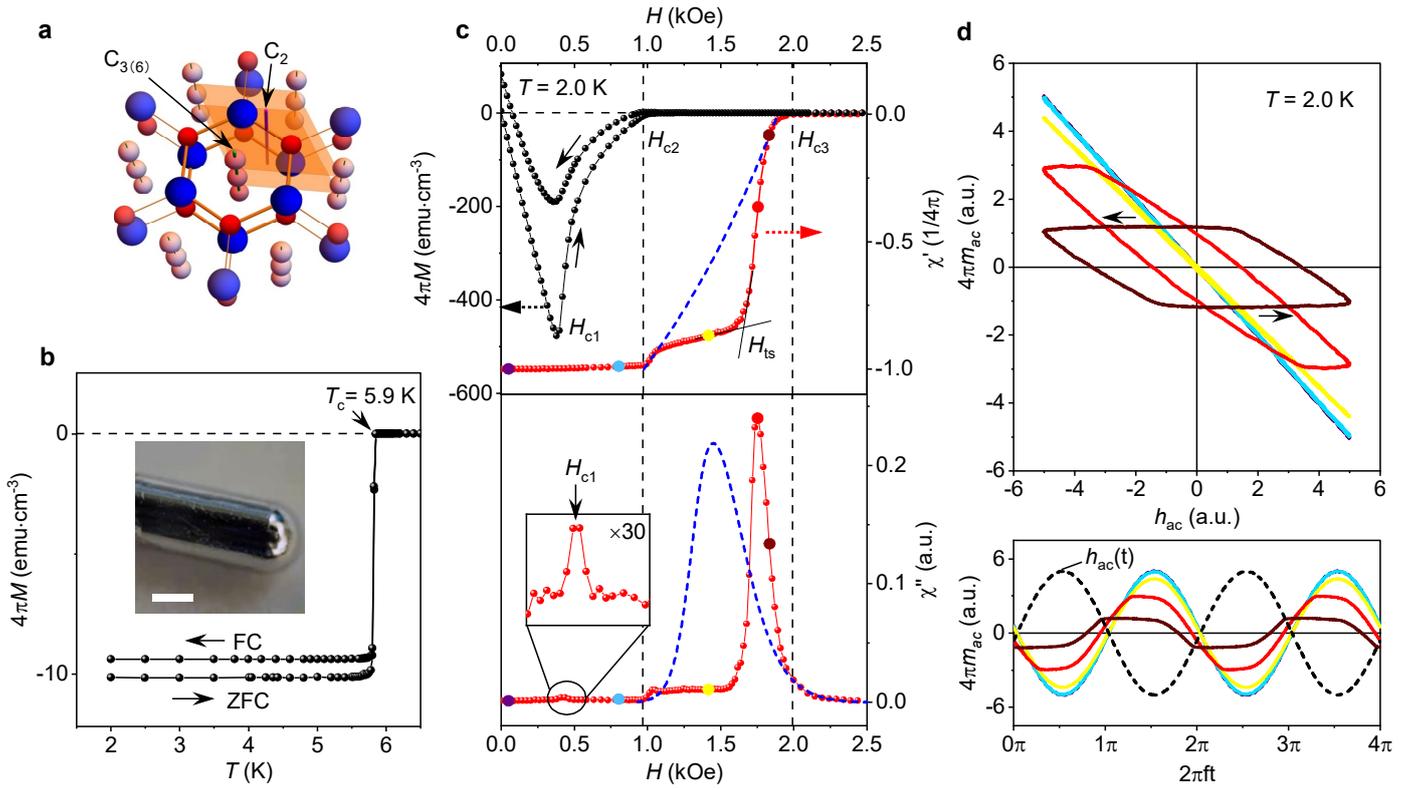

**Figure 1.** Anomalous ac susceptibility of In$_2$Bi. a) Schematic crystal structure of In$_2$Bi. Bi atoms are shown in blue and In atoms in different shades of red, to distinguish between In atoms within the hexagonal planes (dark red) and those making up In chains (light red). The shaded areas denote the unit cell containing four In and two Bi atoms. Symmetry axes are indicated by arrows. b) ZFC and FC magnetization as a function of $T$ at $H = 10$ Oe. Inset: photo of our typical cylindrical crystal; scale bar, 1mm. c) ac susceptibility measured using $h_{ac} = 0.1$ Oe and frequency $f = 8$ Hz (red curves). Black curves: dc magnetization and its hysteresis for this sample. As a reference, the blue dashed curves show the standard response expected for surface superconductivity. The inset in the lower panel shows a zoom of $\chi''$ indicating the transition to the vortex state at $H_{c1}$. The vertical dashed lines indicate $H_{c2}$ and $H_{c3}$, and the arrows the sweep directions. d) Top panel: Lissajous loops for the representative dc fields indicated by the colour-coded dots in (c). Bottom panel: Corresponding waveforms $m_{ac}(t)$ for the applied sinusoidal field $h_{ac}(t)$ of amplitude 0.1 Oe.

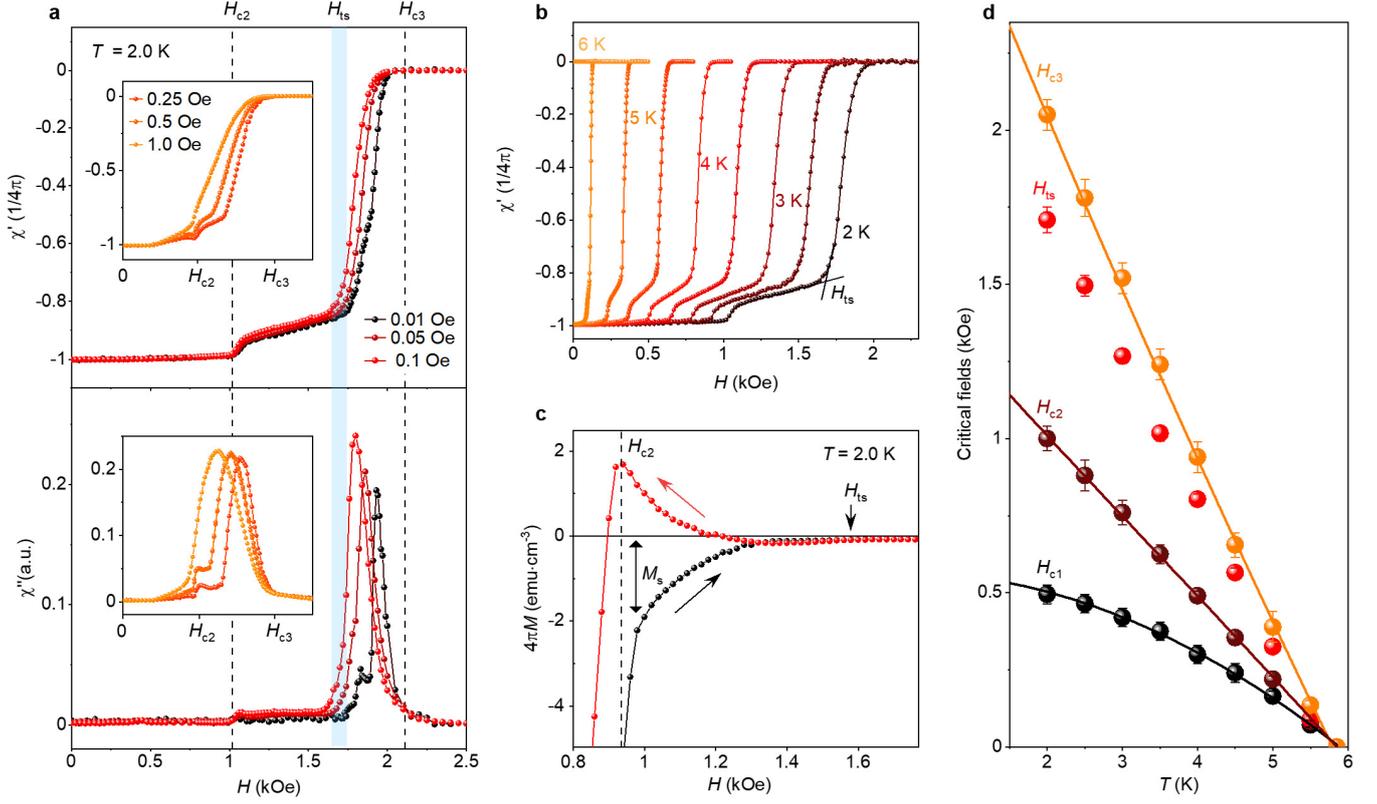

**Figure 2.** Anomalous diamagnetic response at different temperatures and ac excitations. a) ac susceptibility as a function of the ac field amplitude (see legends). b) $\chi'(H)$ at $T$ between 2 and 6 K measured with 0.5 K step; $h_{ac} = 0.1$ Oe. c) Hysteresis in $M(H)$ between the increasing (black symbols) and decreasing (red) dc field $H$; $T = 2$K, $H_{ts}$ is indicated by an arrow. d) Phase diagram for all the critical fields (labelled and colour coded). Red symbols: $H_{ts}(T)$ found from ac susceptibility measurements in (b). Error bars: standard deviations. The black curve shows the standard BCS dependence $H_{c1}(T) \propto 1-(T/T_c)^2$. Brown curve: best fit to $H_{c2}(T)$ using the two-band model of superconductivity (Supporting Information). Yellow curve: guide to the eye. The $H_{c3}/H_{c2}$ ratio changes from 2.0 at 2K to 1.7 at 5.6K, as expected (see text).

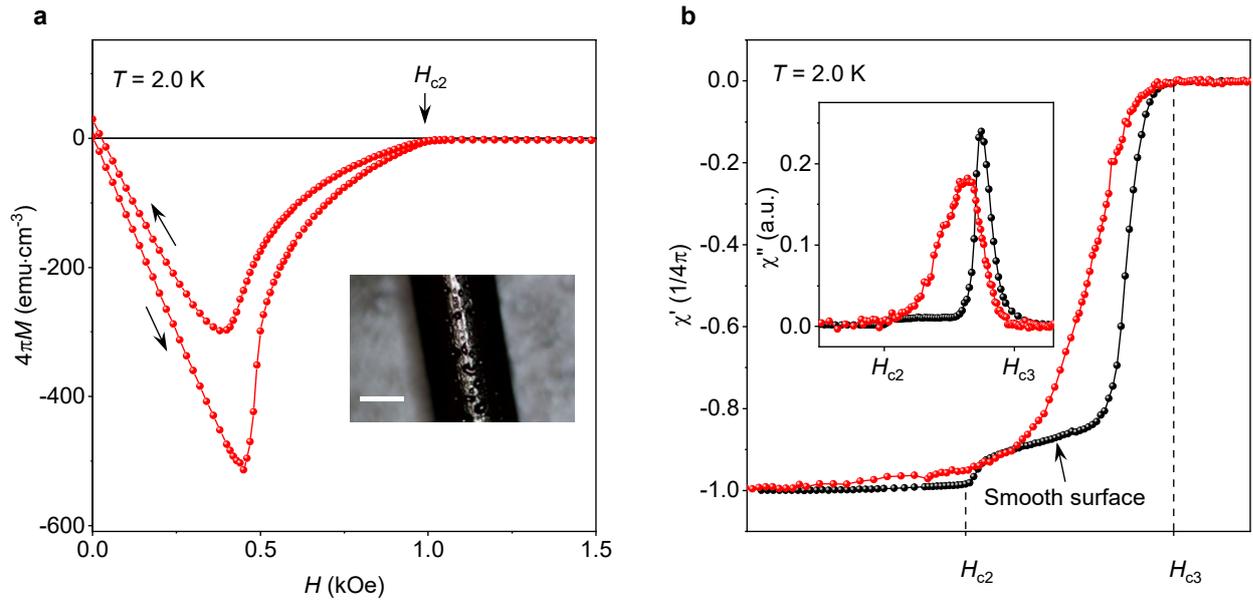

**Figure 3.** Effect of surface quality. a) dc magnetization for a sample with a rough surface shown in the photo (scale bar, 0.5mm). Hysteresis in $M(H)$ between increasing and decreasing field remains small, comparable to our best crystals (cf. Figure 1c). This indicates that surface roughness did not affect quality of the bulk. b) Comparison of ac susceptibility for crystals with comparable bulk pinning but smooth and rough surfaces (black and red curves, respectively). In both cases, $h_{ac} = 0.1$ Oe.

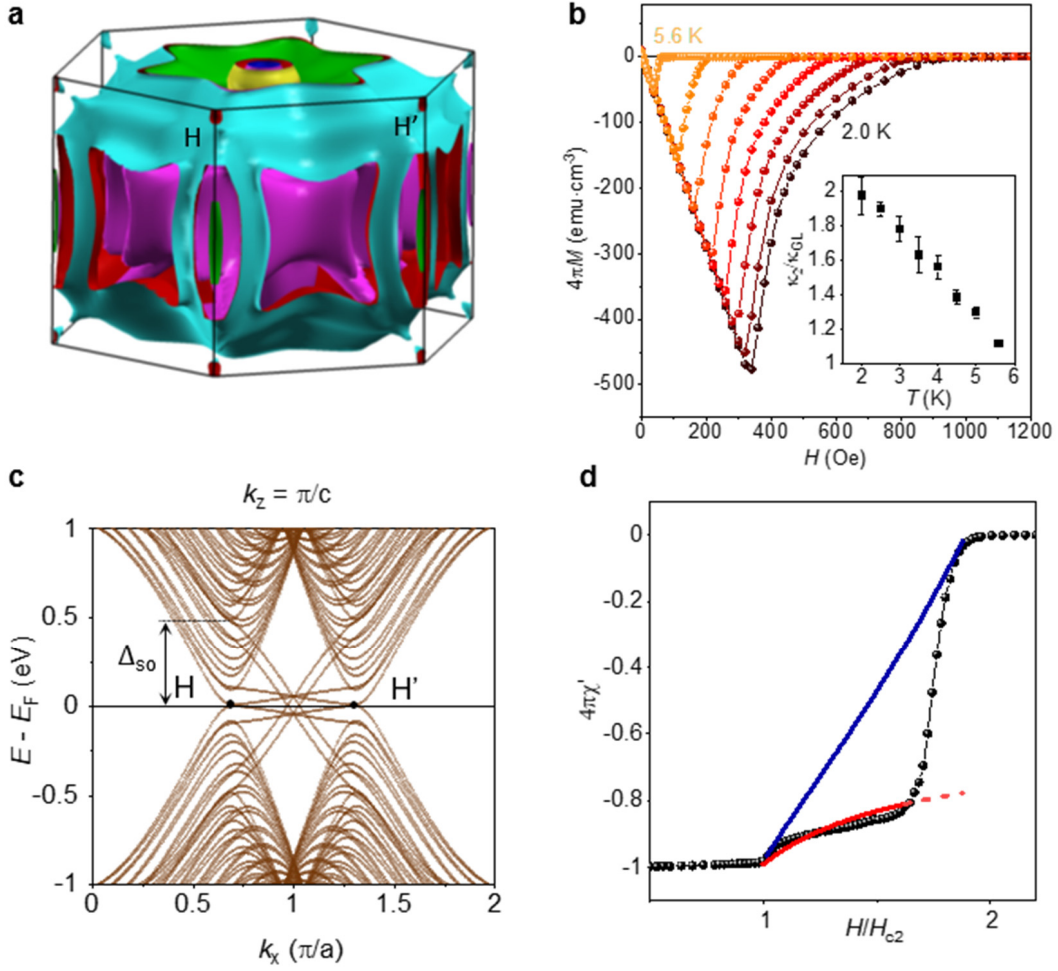

**Figure 4.** Band structure of In$_2$Bi and experimental evidence of multi-band superconductivity. a) Calculated Fermi surface of In$_2$Bi. b) Temperature evolution of dc magnetization. The curves are for $T$ = 2.0, 2.5, 3.0, 3.5, 4.0, 4.5, 5.0 and 5.6 K. Inset: Temperature dependence of the extracted Maki parameter $\kappa_2$. Error bars: standard deviations. c) Band structure of an In$_2$Bi ribbon near H (H') points ('Topological surface states' in Supporting Information). Bands due to In chains are omitted for clarity. Four pairs of counter-propagating edge states cross within the bulk bandgap. Two of the pairs connect bands split by the spin-orbit gap ($\Delta_{SO}$) between Bi-derived bands, while the other two are within a smaller gap of In-derived bands. In these calculations, hopping amplitudes that break particle-hole symmetry were not included. d) Comparison of the observed ac response (symbols) with the theory for conventional surface superconductivity (blue curve) and our model that includes proximitized surface states (red).

**Methods**

*Crystal growth and characterisation.* To grow single crystals of In$_2$Bi, we followed the approach of ref. [44] which is known to result in spontaneous formation of spherical single crystals of 1-2 mm diameter. To this end, Indium (99.99% Kurt Lesker) and Bismuth (99.999% Kurt Lesker) pellets were mixed in stoichiometric composition in a quartz ampoule. The ampoule was sealed and annealed under vacuum (10$^{-6}$ mbar) at 500°C for 24 h. The resulting alloy was re-melted at 150°C in an oxygen- and moisture-free atmosphere of an argon-filled glove box under slow rotation at 1-2 rpm for further homogenisation. This resulted in spontaneous formation of spherical single crystals of $\sim 0.3 - 2$ mm diameter, as reported previously.[44] Following the method of ref. [44], the crystals were kept at 100°C for further 5 mins and then allowed to cool down naturally to room temperature. To grow crystals in a long cylinder geometry, several spherical single crystals were re-melted in a sealed quartz tube of ~2 mm diameter and annealed for two weeks under vacuum at 87°C (just below the melting temperature of In$_2$Bi, 89°C). This produced high-quality cylindrical crystals with smooth surfaces, such as shown in the inset of Figure 1b. All the above procedures and further handling of the crystals were carried out in the protective atmosphere of an Ar filled glovebox (O$_2$ < 0.5 ppm, H$_2$O < 0.5 ppm). Once grown, care was taken to avoid exposure of the crystals to air or moisture by immediately transferring them in the vacuum environment of a cryostat or immersing in paraffin oil. This was necessary to prevent oxidation of Bi at the surface, as we found that a prolonged (few hours) exposure to ambient atmosphere led to formation of a thin surface layer of InBi and/or In$_5$Bi$_3$ (see 'Evidence of In$_2$Bi oxidation in air and the importance of surface protection', Supporting Information).

The monocrystallinity of the samples was confirmed by X-ray diffraction that showed sharp diffraction patterns corresponding to a primitive hexagonal unit cell with $a$ = 5.4728(8) Å and $c$ = 6.5333(12) Å, in agreement with literature for stoichiometric In$_2$Bi. Data on spherical In$_2$Bi crystals were collected in a Rigaku FR-X DW diffractometer using MoKα radiation (λ = 0.71073 Å) at $T$ =150 K and processed using Rigaku CrysAlisPro software.[45] Due to absorption of the diffracted beam by heavy Bi and In atoms, even the 0.3 mm diameter crystal (used to obtain XRD data in Figure S1a) was still too large to collect diffraction data from the whole sample. To overcome this problem, a glancing beam going through different edges of the spherical crystal was used. First the top of the sphere was centred in the beam, then a 4-circle AFC-11 goniometer used to access a wide range of crystal orientations, with the centre of rotation kept at the intersection between the beam and the crystal. Reorienting the crystal allowed us to collect all reflections that fulfil the Bragg condition. See 'Structural characterization of In$_2$Bi crystals' in Supporting Information for further details.

*Magnetization measurements.* Magnetization measurements were carried out using a commercial SQUID magnetometer MPMS XL7 (Quantum Design). Prior to being placed in the magnetometer, samples were mounted inside a low-magnetic background gelatine capsules or a quartz tube, taking care to protect them from exposure to air. In zero-field-cooling (ZFC) mode of dc magnetisation measurements the sample was first cooled down to the lowest available temperature (1.8 K) in zero magnetic field, then a finite field applied and magnetisation measured as a function of an increasing temperature $T$. In field-cooling (FC) mode, the field $H$ was applied above $T_c$ (typically at 10-15 K) and magnetisation measured as a function of decreasing $T$. All ac susceptibility data were acquired with the ac field parallel to the dc field at an excitation amplitude $h_0$ from 0.01 to 2 Oe and a frequency of 8Hz. Test measurements of ac susceptibility at frequencies between 1 and 800 Hz showed that the results were independent of frequency. The superconducting fraction was found as $f = (1 - N) 4\pi |dM/dH|/V$, where $N$ is the demagnetisation factor and $V$ the sample's volume. This yielded $f = 1$, i.e., all our crystals were 100% superconducting.

The superconducting coherence length $\xi$ and magnetic field penetration depth λ were found from the measured critical fields $H_{c1}$ and $H_{c2}$ using the standard expressions [46] $H_{c2} = \Phi_0/2\pi\xi^2$ and $H_{c1} =$ $\left(\Phi_0/4\pi\lambda^2\right)[\ln\kappa + \alpha(\kappa)]$, where $\alpha(\kappa) = 0.5 + (1 + \ln 2)/(2\kappa - \sqrt{2} + 2)$. The Ginzburg-Landau (GL) parameter $\kappa$ was evaluated at all measurement temperatures which showed that it reduced from $\kappa(2K = 0.3T_c) = 1.1$ to $\kappa(T_c) = 0.75$. The critical field for surface superconductivity $H_{c3}$ was determined from ac susceptibility curves such as shown in Figure 1c and Figure S3a. It was defined as the field $H$ corresponding to 0.5% of the χ' value in the Meissner state. For all our crystals we obtained $H_{c3} = 2H_{c2}$ at the lowest measurement temperature, $T = 2K$ (Figure 2d), in agreement with theory for clean superconductors.[47] At higher temperatures, the $H_{c3}/H_{c2}$ ratio gradually decreased, approaching 1.69 close to $T_c$, again in agreement with expectations.[32]

# SUPPORTING INFORMATION

**1. Structural characterization of In$_2$Bi crystals.**

Prior to collecting the Bragg reflections as described in Methods, a pre-experiment was performed to determine the unit cell and orientation matrix for the crystal. During the pre-experiment, reflections were collected and indexed for a range of crystal orientations, giving both the unit cell of the crystal and the orientation matrix that relates the unit cell axes to the instrument axes. Sharp reflections were observed, indicative of a single crystal (Figure S1a). The observed reflections from the pre-experiment data were indexed to a unit cell with a primitive hexagonal Bravais lattice $a$ = 5.471(6) Å, $c$ = 6.515(16) Å, $V$=168.9(5) Å$^3$, indexing against 62 out of 64 observed peaks. This was then used to collect 100% of the unique reflections that were re-indexed to give a primitive hexagonal unit cell of $a$ = 5.4728(8) Å, $c$ = 6.5333(12) Å, V=169.47(5) Å, in excellent agreement with the unit cell reported for In$_2$Bi in literature [1]: $a$ = 5.4760 Å, $c$ = 6.5400 Å, V=169.84 Å$^3$ (at $T = 195$ K).

Additionally, several single crystals were mechanically flattened to turn them into polycrystals and checked for possible presence of a second phase in powder diffraction mode. Typical spectra are shown in Figure S1b and c. The database search was performed against the extracted diffraction patterns using Panalytical X'pert HighScore Plus to index the peaks. This showed excellent correlation with literature for In$_2$Bi (ref. [2]). No other phases could be detected.

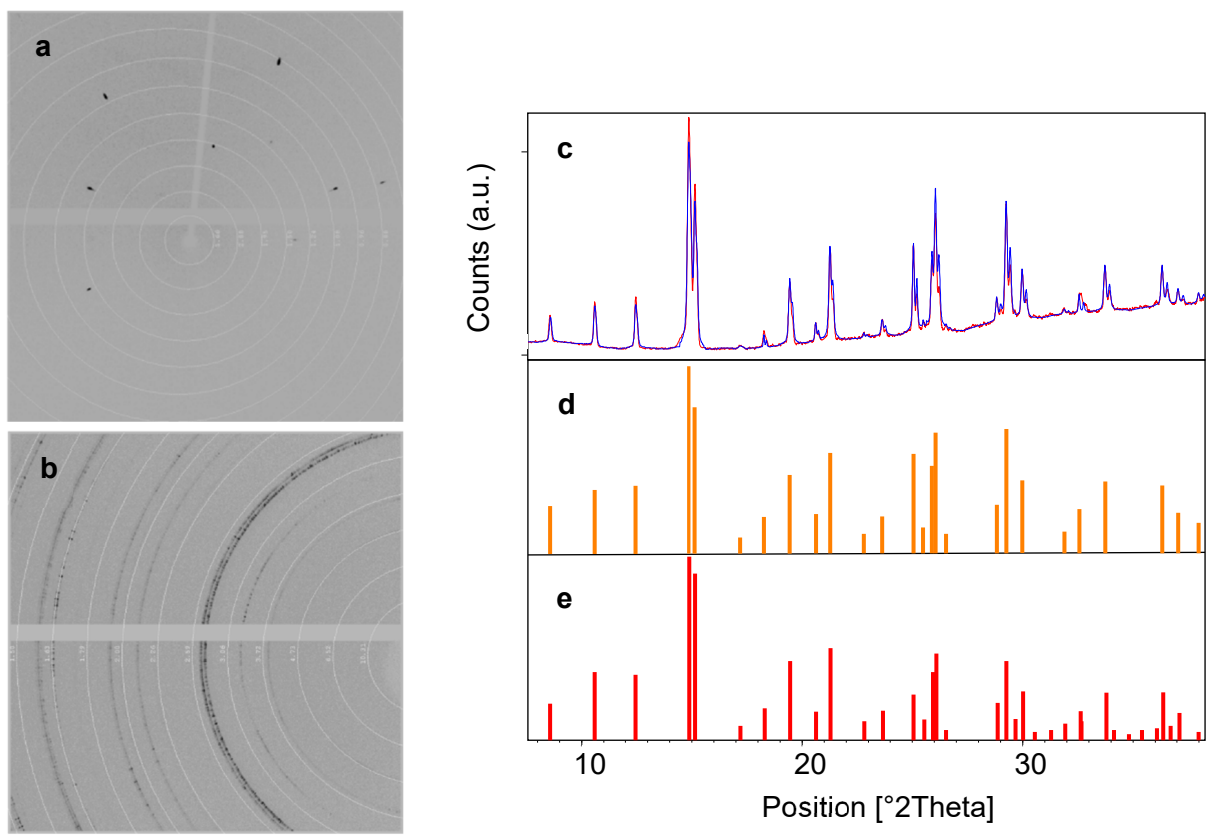

**Figure S1 | Structural characterization of our In$_2$Bi crystals. a,b,** Representative frames from the collected XRD data for an as-grown spherical In$_2$Bi single crystal (**a**) and a polycrystal (mechanically flattened spherical crystal) (**b**). Sharp spots in (**a**) correspond to Bragg reflections for a given orientation of the crystal. Shown are the resolution arcs in Å determined from the distance to the detector and the wavelength used. **c,** Comparison of a measured powder diffraction spectrum for our In$_2$Bi

polycrystal (red line) with a calculated spectrum (blue), indexed and fitted against the database peak positions for In$_2$Bi. **d,e,** Stick representation on the peak positions and relative intensities of the peaks comparing our collected data (**d**) and published data (**e**).

## 2. Evidence of In$_2$Bi oxidation in air and the importance of surface protection.

All magnetisation data in the main text and structural data above were obtained on crystals grown in high vacuum and handled either in the inert (argon) atmosphere of a glove box or immersed in paraffin oil (the latter is known to prevent exposure to oxygen and moisture). This was necessary because an exposure to ambient atmosphere resulted in the appearance of new phases (InBi and In$_5$Bi$_3$) at the surface of the crystals. The presence of In$_5$Bi$_3$ and small amounts of InBi is evident from XRD spectra for samples exposed to air (Fig. S2a) and from the appearance of a second superconducting phase with $T_c \approx 4.2$K in magnetization measurements (Fig. S2b). The above $T_c$ corresponds to the known superconducting transition for In$_5$Bi$_3$ (ref. [3]). The fact that second-phase peaks in XRD spectra are relatively high in intensity compared to the In$_2$Bi host is due to the small penetration depth for Cu-K$\alpha$ X rays, of the order of few μm. From the ratio of the diamagnetic signals corresponding to the superconducting transitions for In$_5$Bi$_3$ and In$_2$Bi, we estimate the thickness of the In$_5$Bi$_3$ layer formed at the surface of In$_2$Bi after several weeks of exposure to air to be around 2 μm (InBi is not superconducting under ambient pressure and therefore does not show up in magnetization measurements).

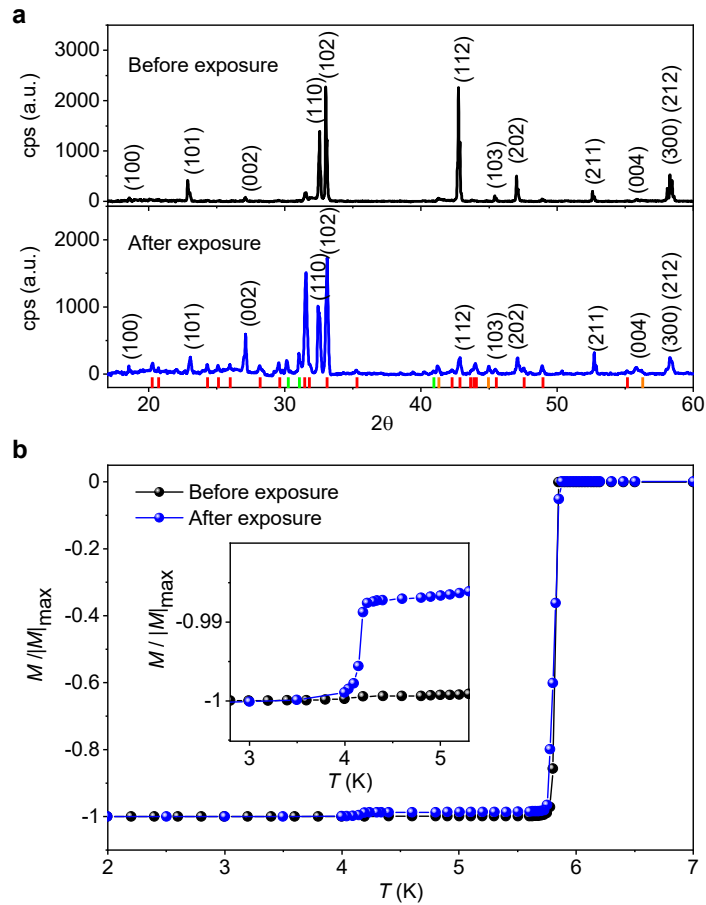

**Figure S2 | Importance of surface protection and evidence of In$_2$Bi oxidation in air. a,** XRD spectra before and after exposure to air for 2 days. All peaks corresponding to In$_2$Bi are labelled. No second phases could be detected in crystals kept in vacuum or in moisture- and oxygen-free environment of a

glovebox. Red, green and orange markers correspond to peak positions for $In_5Bi_3$ (red), InBi (green) or to both phases (orange). XRD data were collected using Rigaku Smartlab diffractometer with Cu Kα radiation ($\lambda$=1.5418 Å). **B**, Normalised $T$-dependent magnetization of an $In_2Bi$ crystal before (black) and after (blue) exposure to air for several weeks. The inset shows a zoomed-up part of the $M(T)$ curves around the expected superconducting transition for $In_5Bi_3$ ($T_c \approx 4.2K$).

A likely reason for the observed formation of $In_5Bi_3$ and InBi is the different enthalpies of oxidation for In and Bi: at room temperature the enthalpy of oxide formation for Bi is $H_{Bi2O3}$ = -575 kJ/mol and for In it is $H_{In2O3}$ = -924 kJ/mol, that is, indium is oxidized more easily. In turn, formation of $In_2O_3$ at the surface leads to In deficiency, favouring formation of $In_5Bi_3$ or/and InBi. The $In_2O_3$ grown at the surface is likely to be amorphous and therefore does not produce any peaks in XRD spectra.

## 3. Magnetization of spherical single crystals and effect of bulk pinning.

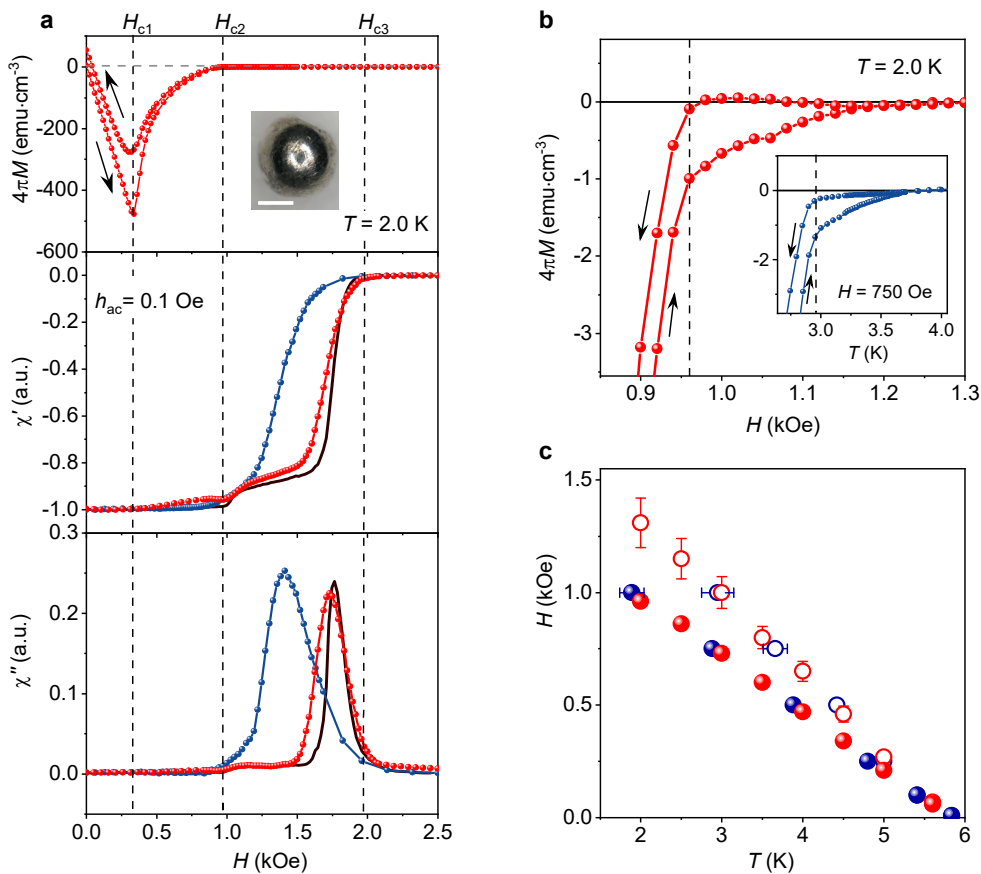

**Figure S3| Magnetic response of spherical $In_2Bi$ crystals**. **a,** Typical field-dependent dc magnetisation $M(H)$ of $In_2Bi$ spheres (top panel) and corresponding ac susceptibility (middle and bottom panels, red symbols). Shown are measurements at temperature $T = 2$ K and ac field amplitude $h_0 = 0.1$ Oe. For comparison, also shown are data for the cylindrical sample of Fig. 1c in the main text (black lines) and for a spherical crystal with intentionally degraded (sandpapered) surface (blue symbols). The inset in the top panel shows a photo of the crystal; scale bar 1 mm. **b,** *Main panel*: Hysteresis in field-dependent dc magnetisation, $M(H)$, above the bulk transition to the normal state; temperature $T = 2K$. Up/down arrows indicate measurements in increasing /decreasing external field. *Inset*: Hysteresis in temperature-dependent magnetisation $M(T)$ measured at $H = 750$ Oe. Up/down arrows correspond to zero-field cooling (ZFC)/field cooling (FC), respectively. Vertical dashed lines indicate the field $H$ or temperature $T$ corresponding to the bulk transition to the normal state. **c,** Phase diagram for the $In_2Bi$ sphere from **b.**

Solid red symbols correspond to the bulk transition to the normal state, $H_{c2}(T)$, and solid blue symbols to $T_c(H)$. Open symbols correspond to the disappearance of the hysteresis in $M(H)$ (red) and in $M(T)$, blue. The agreement between the two types of measurements validates the common origin of the diamagnetic surface contribution.

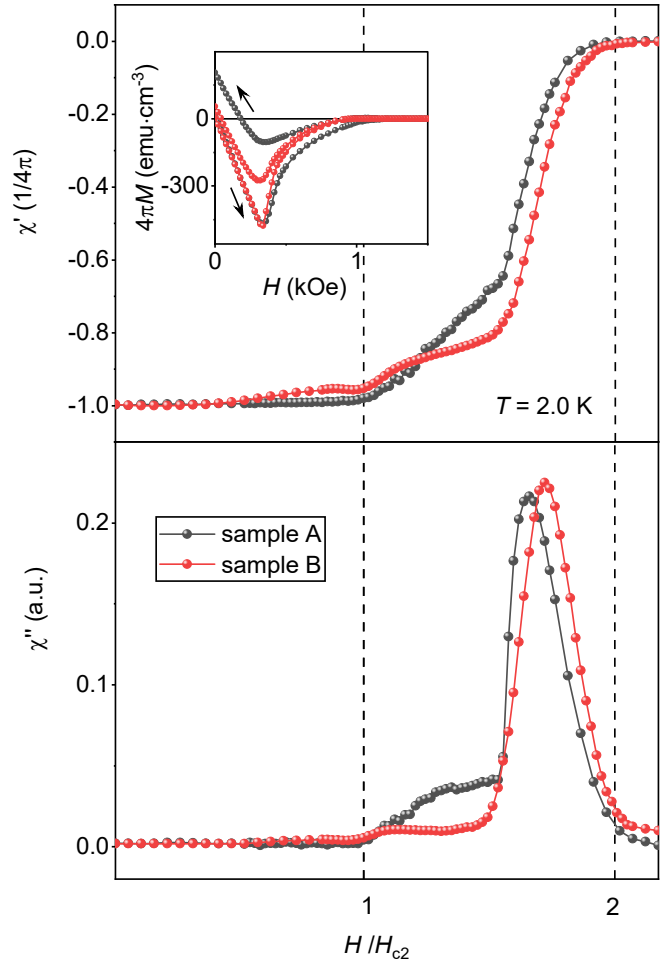

**Figure S4 | Effect of bulk pinning on the diamagnetic response of the surface sheath and the transition at $H_{ts}$.** Shown are data for two spherical crystals with different bulk pinning strengths. Stronger bulk pinning for sample A is indicated by a larger hysteresis in dc magnetization $M(H)$ between increasing and decreasing $H$ (indicated by arrows in the inset of the top panel) and a larger remnant $M$ at zero $H$. The transition at $H_{ts}$ in ac susceptibility is clearly visible but the diamagnetic screening below $H_{ts}$ is weaker compared to sample B (data for sample B also shown in Fig. S3). Bulk pinning is an indication of non-uniformities and the presence of defects in a crystal which typically results in local variations of the superconducting coherence length. To some extent it can be expected to affect the near-surface of the crystal, too, weakening its diamagnetic response [25].

## 4. Temperature dependence of $H_{c2}$: fitting to the multiband theory.

To analyse the experimental temperature dependence of $H_{c2}$ we use the two-band model proposed by Gurevich *et al* [4,5]. As discussed in literature [6], the more sensitive indicator of the multiband nature of superconductivity is a strong temperature dependence of the slope of dc magnetisation $\frac{dM}{dH}|_{H=H_{c2}}$, as we indeed observe for In$_2$Bi. Contributions from multiple bands also modify the temperature dependence of $H_{c2}$. The model [4,5] takes into account multiple scattering channels that are included via intraband- and interband electron-phonon coupling parameters, $\lambda_{11}$, $\lambda_{22}$ and $\lambda_{21}$, $\lambda_{12}$, respectively, and normal state electronic diffusivity tensors, $D_m^{\alpha\beta}$, reflecting the underlying symmetry and anisotropy of the Fermi surfaces [4]. An anomalous $T$ dependence of $H_{c2}$ (enhancement at low $T$) results from different diffusivities for different electronic bands: In refs. [4-6] the model was compared with the well-known example of two-band superconductivity in MgB$_2$ where the principal diffusivity value $D_\sigma^{(c)}$ along the c axis is much smaller than the two in-plane values $D_\sigma^{(a)}$ and $D_\sigma^{(b)}$ due to the nearly 2D nature of the σ band (band 1). In contrast, for the 3D π band (band 2), the difference in principal values $D_\pi^{(a)}$, $D_\pi^{(b)}$, and $D_\pi^{(c)}$ is less pronounced, resulting in a disparity of intraband diffusivities $D_1$ and $D_2$. From the band structure and Fermi surface topology (Figs 4a, S6, S8), the situation is similar for our In$_2$Bi where the diffusivity for the electronic states associated with hexagonal In$_1$Bi$_1$ planes can be expected to be different from that for the electronic states having a 3D character. This qualitative picture is born out in the observed strong temperature dependence of the slope of dc magnetisation $\frac{dM}{dH}|_{H=H_{c2}}$ described by the Maki parameter $\kappa_2$ (Fig. 4b) and its ratio to the GL parameter $\kappa_{GL} = \lambda/\xi$. It follows from our measurements that both $\kappa_{GL}$ and $\kappa_2$ for In$_2$Bi are temperature dependent, with $\kappa_2 = 0.75 = \kappa_{GL}$ near $T_c$, as expected [6], and their ratio increasing to $\kappa_2/\kappa_{GL} \approx 2$ at our lowest measurement temperature, 2K (inset in Fig. 4b in the main text). According to analysis of ref. [6], such a large increase corresponds to a diffusivity ratio for different bands $\eta \sim 0.1$, in agreement with the best fit to our experimental $H_{c2}(T)$ (see Fig. S5) obtained using an implicit expression [4,5]

$$a_0[\ln(t) + U(h)][\ln(t) + U(\eta h)] + a_1[\ln(t) + U(h)] + a_2[\ln(t) + U(\eta h)] = 0 \quad (1)$$

where $U(x) = \psi(x + 1/2) - \psi(1/2)$, $\psi(x)$ is the digamma function; $a_0 = 2w/\lambda_0$, $a_1 = 1 + \lambda_-/\lambda_0$, $a_2 = 1 - \lambda_-/\lambda_0$, $\lambda_- = \lambda_{11} - \lambda_{22}$, $w = \lambda_{11}\lambda_{22} - \lambda_{12}\lambda_{21}$, $\lambda_0 = (\lambda_-^2 + 4\lambda_{12}\lambda_{21})^{1/2}$, $\eta = D_2/D_1$, $t = T/T_c$, and $h = H_{c2}D_1/2\phi_0 T$. To obtain the best fit, we set the diffusivity ratio $\eta$ as a fitting parameter and tested different sets of coupling parameters $\lambda$. This showed that the fit is very sensitive to $\eta$, with the best fit corresponding to $\eta = 0.1$ (Fig. S5), i.e., the same value as inferred from the $T$ dependence of $M(H)$ at $H_{c2}$. In contrast, the fit was found to be practically insensitive to the coupling constants $\lambda_{ii}$ and $\lambda_{ij}$ in the available temperature range (inset in Fig. S5) indicating that $H_{c2}$ data alone are insufficient to derive information about electron-phonon couplings in different bands. Importantly, this has no bearing on our discussion and/or conclusions on the role of the topological surface states.

As expected [4-6], the extrapolated value of $H_{c2}(0) \approx$ 1.5 kOe (Fig. S5) is considerably higher than the universal value for a single-band superconductor with $H_{c2}$ limited by orbital pair breaking [7], $H_{c2}^{orb}(0) \approx -0.693 T_c (\frac{dH_{c2}}{dT})|_{T=T_c}$. For our crystals with $T_c$ = 5.9 K and $\frac{dH_{c2}}{dT}|_{T=T_c} \approx$ 220 Oe/K, $H_{c2}^{orb}(0) \approx$ 0.9 kOe.

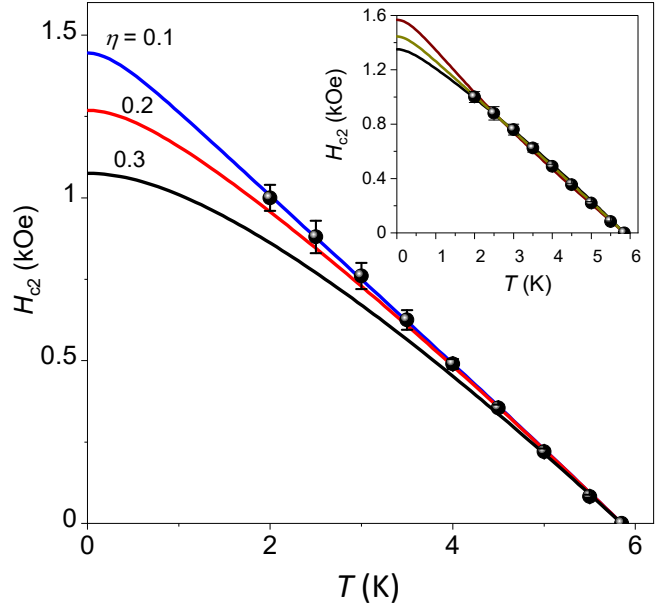

**Figure S5 | Temperature dependence of the upper critical field.** Shown are $H_{c2}(T)$ data for the cylindrical single crystal of Fig. 2d in the main text. *Main panel:* Solid lines: fits to eq. (1) for the same set of $\lambda_{ij}$ ($\lambda_{11} = 0.7$; $\lambda_{22} = 0.3$; $\lambda_{12} = \lambda_{21} = 0.8$) and different diffusivity ratios shown as labels. *Inset:* Fits to eq. (1) for $\eta = 0.1$ and three different sets of $\lambda_{ij}$. Top curve: $\lambda_{11} = 0.6$; $\lambda_{22} = 0.4$; $\lambda_{12} = \lambda_{21} = 0.6$; middle curve: $\lambda_{11} = 0.8$; $\lambda_{22} = 0.2$; $\lambda_{12} = \lambda_{21} = 0.9$; bottom curve: $\lambda_{11} = 0.7$; $\lambda_{22} = 0.3$; $\lambda_{12} = \lambda_{21} = 0.8$. In the temperature range where data are available, the fit is equally good for all sets of $\lambda_{ij}$. Fits to $H_{c2}(T)$ for all our crystals (cylindrical and spherical) produced similar results.

## 5. DFT analysis

Our first-principles electronic structure calculations are based on density functional theory (DFT) [8,9] as implemented in the Vienna *ab initio* simulation package (VASP) [10,11] and a generalized gradient approximation of Perdew-Burke-Ernzerhof-type [12] is employed for the exchange-correlation energy. The electron-ion interaction is described by the projector augmented-wave (PAW) method [13] and the *d* semi-core states are included in the In and Bi PAW datasets used. The spin-orbit coupling is included in all calculations. Wave functions are expanded in terms of plane waves with a kinetic energy cutoff of 400 eV. The ground-state charge density is evaluated on a 32 x 32 x 24 *k*-point mesh. The plane-wave cut-offs and *k*-point meshes are chosen to ensure the convergence of total energies within 0.5 meV. The *ab initio* tight-binding Hamiltonian is constructed using maximally-localized Wannier functions [14] as a basis, and *s* and *p* orbitals centred on In and Bi atoms are used as projection orbitals. In order to preserve the symmetry of the resulting Wannier functions as much as possible, an iterative minimization step was avoided. The resulting Hamiltonian was then used to calculate momentum-resolved density of states using the iterative Green's function method [15]. The DFT results, highlighting the contributions from Bi and In p-orbitals, are shown in Fig. S6.

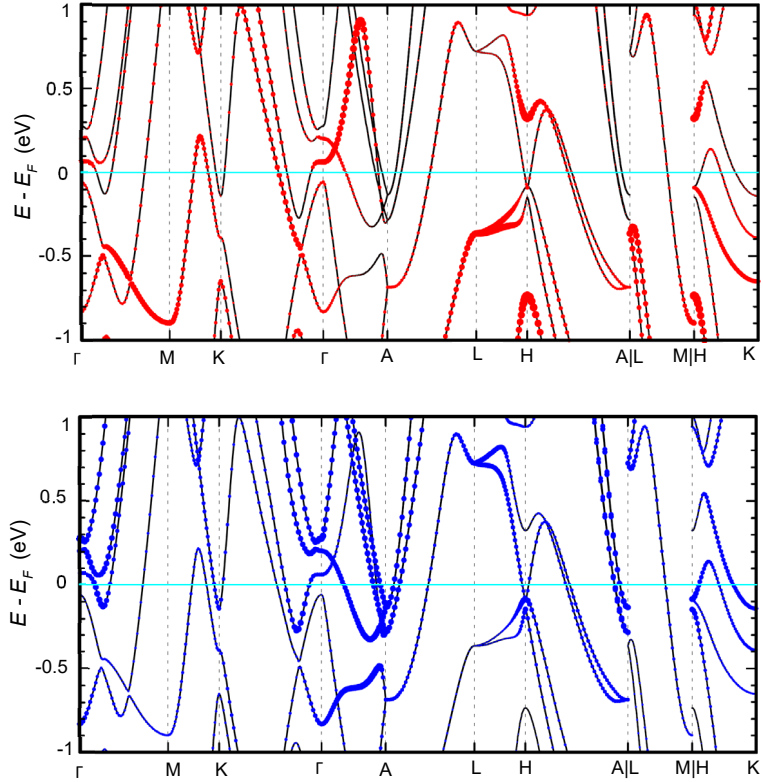

**Figure S6 | Band structure of In$_2$Bi.** Bulk band structure calculated using density functional theory. The size of red (blue) dots indicates *p*-orbital contributions to different bands from Bi (top panel) and In (bottom panel) atoms.

### 6. Tight-binding calculations

Looking at the crystal structure of In$_2$Bi (Fig. 1a of the main text), we recognise In$_1$Bi$_1$ honeycomb planes arranged in an AA' configuration (In atoms on top of Bi atoms, and vice versa) and 1D In chains passing through the hexagon centres. The Bi-Bi and In-In distance in the In$_1$Bi$_1$ planes are assumed to be the same and equal to a parameter $a$. The distance between consecutive In$_1$Bi$_1$ planes, as well as between the In atoms within the 1D chains is given by $c/2$. The symmetric unit cell (shaded region in Fig. 1a) contains six atoms: two In atoms from the vertical chains, and two In and two Bi atoms from In$_1$Bi$_1$ planes. Its height is $c$. The In (Bi) atom of one layer is mapped onto the In (Bi) atom of the other layer by a screw transformation. This comprises a translation by $c/2$ in the vertical direction, which maps the In (Bi) atom of one layer into the Bi (In) of the other, followed by 180° rotation with respect to the vertical axis passing through the midpoint of the cell. The latter transformation maps the In (Bi) atom onto the Bi (In) atom of the same In$_1$Bi$_1$ layer. Both operations separately leave the In-wire subsystem invariant. Therefore, their combination leaves the whole system (In$_1$Bi$_1$ planes and In chains) invariant. The crystal also exhibits a 3-fold rotational (C$_3$) axis passing through the centre of an In$_1$Bi$_1$ hexagon, which also serves as a 6-fold rotational screw-symmetry axis (C$_6$), when combined with the screw symmetry above.

Based on the crystallographic considerations, we can construct a minimal tight-binding model that captures the salient features of the In$_2$Bi band structure and allows us to gain insight into the *ab-initio* DFT results (Fig. S6). As in the DFT calculations, we use the electronic configurations of In and Bi, [Kr]4d$^{10}$5s$^2$5p$^1$ and [Xe]4f$^{14}$5d$^{10}$6s$^2$6p$^3$, respectively. Accordingly, the In electrons that contribute most to the properties of the compound are those in 5s and 5p orbitals and the contribution of Bi atoms is dominated by 6s and 6p electrons. Furthermore, Bi is a heavier element and has a much stronger spin-orbit interaction. To keep our analysis as simple as possible, we include only one p-like orbital per atom. This is sufficient to reproduce the main features of the band structure around the $H$ (and $H'$) point of the Brillouin zone (Fig. S8a). Some details of the DFT calculations (e.g., the 12 bands crossing the Fermi surface) do not appear in the simple model and would require finer details of orbital hybridization to be included in the tight-binding analysis. Such details are not essential for our purpose here, as the simple model is already capable of explaining the occurrence of topological surface states (see the following section 'Topological surface states').

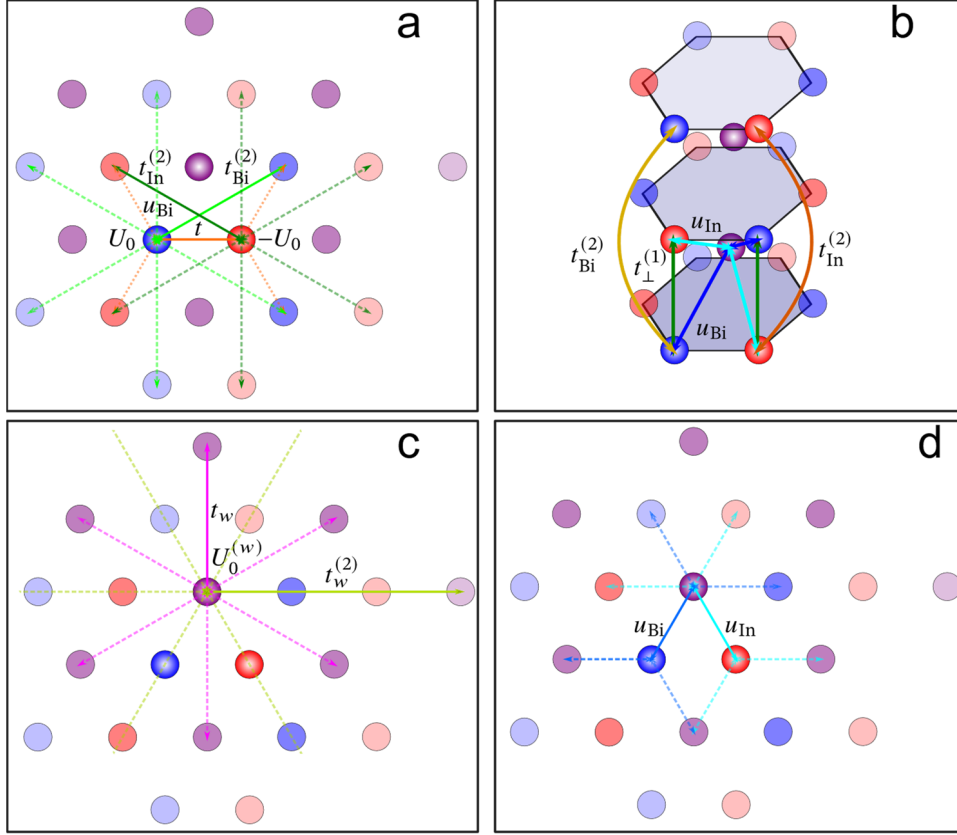

**Figure S7: The structure of our tight-binding model.** Hamiltonians $H_{InBi}$ (**a**), $H_{InBi-In}$ (**b**, **d**) and $H_{In}$ (**c**). The atoms shown in 3D are those belonging to the symmetric unit cell; **a,c,d** are top views, and **b** is a side view. The purple atoms represent the In chains' atoms at height $c/4$ and $3c/4$ from the bottom $In_1Bi_1$ layer of the unit cell. The red (blue) circle stand for In (Bi) atom of the bottom $In_1Bi_1$ layer and a Bi (In) atom of the $In_1Bi_1$ layer at height $c/2$.

Our tight-binding model is shown schematically in Fig. S7. To calculate the electronic bands corresponding to the hexagonal $In_1Bi_1$ planes, we express all parameters in terms of the intralayer hopping between In and Bi atoms, which we call $t$. The on-site energy of a Bi (In) atom is $+U_0$ ($-U_0$). Keeping in mind that relativistic corrections play an important role, we introduce next-nearest-neighbour Kane-Mele spin-orbit couplings $t_\alpha^{(2)}$, where $\alpha$ = Bi, In (because of their atomic weights, the two atoms are expected to exhibit very different spin-orbit interactions). We use $t_\perp^{(1)}$ to denote the interlayer hopping parameter that couples an In (Bi) atom with the Bi (In) directly above. Similarly, $t^{(2)}_{\perp,\alpha}$ denotes the In-In and Bi-Bi interlayer hopping parameter. The Hamiltonian describing In$_1$Bi$_1$ layers is therefore:

$$H_{\text{InBi}} = U_0 \sum_{i,l,\alpha} \sigma^z_{\alpha\alpha} c^+_{il\alpha} c_{il\alpha} + t \sum_{\langle i,j\rangle,l,\alpha,\alpha'} c^+_{il\alpha} \sigma^x_{\alpha\alpha'} c_{jl\alpha'} + i \sum_{\langle\langle i,j\rangle\rangle,l,\alpha} t^{(2)}_\alpha \sigma^z_{\alpha\alpha} s^z_{\alpha\alpha} c^+_{il\alpha} c_{jl\alpha}$$
$$+ t^{(1)}_\perp \sum_{\langle l,l'\rangle,i,\alpha,\alpha'} c^+_{il\alpha} \sigma^x_{\alpha\alpha'} c_{il'\alpha'} + \sum_{\langle\langle il,jl'\rangle\rangle,\alpha} t^{(2)}_{\perp,\alpha} c^+_{il\alpha} c_{jl'\alpha},$$

where $c_{il\alpha}$ ($c^+_{il\alpha}$) destroys (creates) an electron in the $\alpha$ = In, Bi atom at position $i$ of the $l$-th In$_1$Bi$_1$ layer, $\sigma^x_{\alpha\alpha'}, \sigma^y_{\alpha\alpha'}, \sigma^z_{\alpha\alpha'}$ ($s^x_{\alpha\alpha'}, s^y_{\alpha\alpha'}, s^z_{\alpha\alpha'}$) are Pauli matrices acting on the sublattice (spin) degree of freedom, while $\langle i,j\rangle$ and $\langle\langle i,j\rangle\rangle$ denote that the summation is restricted to nearest- and next-nearest-neighbour atoms. Finally, the sum over $\langle\langle il,jl'\rangle\rangle$ runs over next-nearest-neighbour atomic sites on adjacent layers.

The vertical In chains are described with an on-site energy $U_0^{(w)}$ and a nearest-neighbour (next-nearest-neighbour) hopping $t_w$ ($t_w^{(2)}$) in the vertical direction. Their Hamiltonian reads

$$H_{\text{In}} = U_0^{(w)} \sum_{a,\lambda} d^+_{a\lambda} d_{a\lambda} + t_w \sum_{\langle\lambda,\lambda'\rangle,a} d^+_{a\lambda} d_{a\lambda'} + t_w^{(2)} \sum_{\langle\langle\lambda,\lambda'\rangle\rangle,a} d^+_{a\lambda} d_{a\lambda'},$$

where $d_{a\lambda}$ ($d^+_{a\lambda}$) destroys (creates) an electron in the $\lambda$-th atom of the In wire at in-plane position $a$. Finally, we couple the In chains to In$_1$Bi$_1$ planes via the hopping parameters $u_\alpha$ ($\alpha$ = In, Bi).

$$H_{\text{InBi-In}} = \sum_{\langle il\alpha,\lambda a\rangle} u_\alpha (d^+_{a\lambda} c_{il\alpha} + c^+_{il\alpha} d_{a\lambda}).$$

Here $\langle il\alpha, \lambda a\rangle$ restricts the sum to nearest-neighbour atoms belonging to an In$_1$Bi$_1$ hexagon and an In 'chain'. The full Hamiltonian of the system is $H = H_{\text{InBi}} + H_{\text{In}} + H_{\text{InBi-In}}$.

The band structure is obtained by diagonalizing the combined Hamiltonian, $H_{\text{InBi}} + H_{\text{In}} + H_{\text{InBi-In}}$. The eigenvalues have a complicated analytical form that is not reported here. A representative band structure is shown in Fig. S8b. To obtain this result, we have fitted the tight-binding parameters to the DFT results (Fig. S6) aiming to reproduce the band structure at high-symmetry points of the Brillouin zone. We find $t = 0.45$ eV, $U_0 = 0.4$ eV, $t^{(2)}_{Bi} = 0.05$ eV, $t^{(1)}_\perp = -0.65$ eV, $t^{(2)}_{\perp,In} = t^{(2)}_{\perp,Bi} = 0.4$ eV, $U_0^{(w)} = 0.55$ eV, $t_w = 0.6$ eV, $t_w^{(2)} = -0.2$ eV, $u_1 = u_2 = 0.05$ eV. In the resulting spectrum (Fig. S8b) one recognises a series of band crossings along the $\Gamma - A$ line, two-fold degenerate bands in the $A - L$ direction (forming a "nodal" line) and high-density-of-states bands near $\Gamma$ and H (H') points in the Brillouin zone. These features are a direct consequence of the In$_2$Bi nonsymmorphic crystal symmetry. The degeneracy of the Dirac crossings and nodal lines is in fact a consequence of the different transformation properties of the states under C$_2$ screw-symmetry [16,17]. Since the crystal potential respects such symmetries, the coupling between those states must vanish, no gap can be opened and they remain degenerate. Equivalently, because of the C$_2$ screw-symmetry, the vertical coupling between In$_1$Bi$_1$ planes must vanish at $k_z = \pi/c$, so that they represent two copies of an "asymmetric" Kane-Mele model. Such decoupling explains the twofold degeneracy of the bands along the nodal $A - L$ line. In turn, the symmetry-protected band crossings and nodal lines imply the existence of surface states in the normal state of In$_2$Bi [18-20]. Those are discussed in the next section.

Furthermore, the coupling between In$_1$Bi$_1$ planes and In chains is small, too, as can be seen from the values of $u_{1,2}$, suggesting that the two sub-systems contribute nearly independently to the overall response of In$_2$Bi. The electronic states representing In$_1$Bi$_1$ planes and In chains have different dimensionalities. The states at H (H′) localised in In$_1$Bi$_1$ planes have a pure 2D character, while those associated with In chains are more of a 3D character. Superconductivity in such a system can therefore be expected to exhibit a coexistence of two weakly coupled superconducting gaps with different dimensionalities, in agreement with the experimental observations.

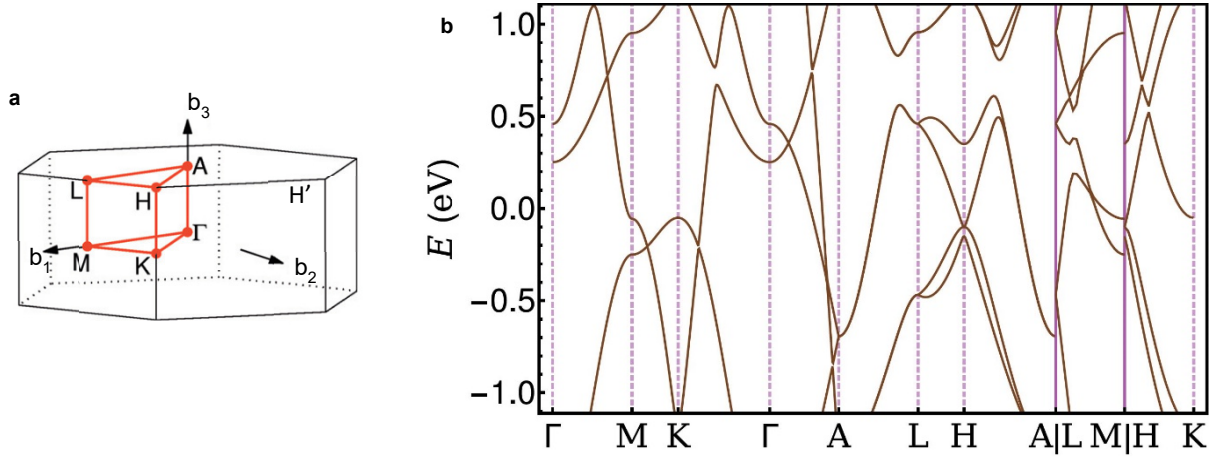

**Figure S8 | Band structure of In$_2$Bi. a,** Brillouin zone of In$_2$Bi. **b,** Bulk band structure calculated using a tight-binding model fitted to *ab-initio* DFT simulations shown in Fig. S6. Positions of high-symmetry points in the Brillouin zone are shown in **a**.

### 7. Topological surface states

Surface states in In$_2$Bi stem from the nontrivial topology of In$_1$Bi$_1$ planes and are protected by the screw symmetry as described in the previous section. To show how such states emerge, we consider a thin film modelled as a stack of In$_1$Bi$_1$ layers, finite in one direction (and terminated with zigzag edges) and infinite in the other. We apply periodic boundary conditions in the latter direction. We further simplify the model introduced in section 6 by neglecting all next-nearest-neighbour hopping amplitudes, with the crucial exception of the Kane-Mele-type spin-orbit couplings that are essential for the emergence of topologically protected surface states. We note that the resulting model lacks some of the features of the full one introduced in section 6 (for example, it does not account for the particle-hole asymmetry of the band structure). Such features, arising from the weak coupling between In$_1$Bi$_1$ planes and In chains, are not important for describing the surface states.

The unit cell of In$_2$Bi (shown in Fig. 1a of the main text) encompasses two consecutive layers and contains two atoms per row $j = 1, \ldots, N$, one in each layer. For calculation purposes, it is convenient to introduce units composed by pairs of rows ($j$ and $j + 1$) each containing four atoms, one In and one Bi per layer. The In and Bi atoms in the two different rows are distinguished by the sublattice degree of freedom. We define $\Gamma_{abc} = \tau^a \otimes \sigma^b \otimes s^c$ where $\tau^a, \sigma^a$ and $s^a$ ($a = x, y, z$) are three sets of Pauli matrices operating on the layer, sublattice and spin degrees of freedom, respectively, while $\otimes$ denotes the tensor product. The Hamiltonian operating on the four sites in two consecutive rows of the unit cell is

$$H_{\tilde{j}\tilde{j}} = 2\sin(k_x a)\left(\frac{t_{Bi}^{(2)} + t_{In}^{(2)}}{2}\Gamma_{333} + \frac{t_{Bi}^{(2)} - t_{In}^{(2)}}{2}\Gamma_{303}\right) - U_0\Gamma_{330} + 2t\cos\left(\frac{k_x a}{2}\right)\Gamma_{010}$$
$$+ t_\perp^{(1)} \cos\left(\frac{k_z c}{2}\right)\Gamma_{100},$$

whereas the hopping between successive pairs of rows is given by

$$H_{\tilde{j}\tilde{j}+1} = -2\sin\left(\frac{k_x a}{2}\right)\left(\frac{t_{Bi}^{(2)} + t_{In}^{(2)}}{2}\Gamma_{333} + \frac{t_{Bi}^{(2)} - t_{In}^{(2)}}{2}\Gamma_{303}\right) + \frac{t}{2}(\Gamma_{010} + i\Gamma_{020}).$$

In these equations, $\tilde{j} = 1, \ldots, N/2$ denotes the pair of rows, while $k_x$ is the quasi-momentum along the in-plane infinite direction. Conversely, $k_z$ is the quasi-momentum in the direction orthogonal to the $In_1Bi_1$ planes. The other parameters entering these equations and their numerical values are given in section 6.

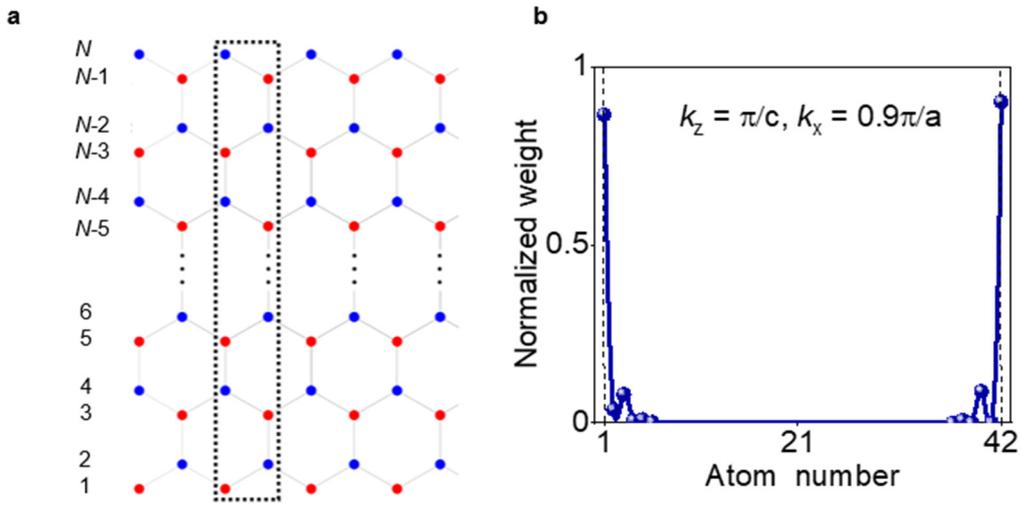

**Figure S9 | Spatial localization of topological surface states. a,** Schematic of the lattice used in our finite-size calculations. The crystal is modelled as an infinite stack of $In_1Bi_1$ hexagonal sheets, with In (Bi) aligned on top of Bi (In). Each sheet is terminated with zigzag edges and is infinite in the other direction. The unit cell encompasses two consecutive layers and contains N atoms in each of them (along the finite direction). The dashed line marks the unit cell in one such layer. Periodic boundary conditions are applied in both the in-plane infinite direction, and in the direction perpendicular to the planes. **b,** Normalised weight of the edge state wavefunction on the atoms of one layer of the InBi unit cell (symbols). The edge state has exactly the same weight on the atoms of the other layer. In this calculation, the number of atoms in each layer *N*=42. Connecting line is a guide to the eye.

The full Hamiltonian of the film is obtained by combining $N/2$ $H_{\tilde{j}\tilde{j}}$-blocks along the diagonal of an $4N \times 4N$ matrix. Different pairs of rows are connected by off-diagonal $H_{\tilde{j}\tilde{j}+1}$-blocks. The resulting matrix is diagonalized for fixed $k_z$ and as a function of $k_x$. Representative results are shown in Fig. 4c in the main text: At $k_z = \pi/c$ two Dirac cones appear from the projection of the bulk band structure onto the surface Brillouin zone. Edge states connect bulk bands and cross the gap, forming a series of linear crossings. This in turn implies that surface states cross the Fermi energy which, according to DFT calculations (section 5) is located around the Dirac points of the bulk band structure. Finally, the weights of the edge states on the atoms of a given layer (Fig. S9b) are obtained by projecting the eigenstates of the $4N \times 4N$ Hamiltonian into given spin and layer components.

# 8. AC susceptibility and dc magnetization in conventional superconductors: Contribution of surface superconductivity

Magnetization and susceptibility measurements on superconductors detect signals that have their origins in circulating persistent shielding currents [21-26]. The basic idea is that a superconducting surface sheath can support a finite (non-zero) current; as long as this sheath of current is continuous, it will screen the total volume of a superconductor, irrespective of whether its bulk is in normal or mixed state [23-26].

For a superconducting cylinder in a parallel dc magnetic field $H_a$ and a superimposed ac field $h = h_0 e^{i\omega t}$ ($h \parallel H_a$), the susceptibility is given by [21,22]

$$\chi = \frac{1}{4\pi}\left[-1 + \frac{2}{d^2 H_a}\int_0^d B(r) r\, dr\right]$$

where $B(r)$ is the magnetic induction inside the sample and $d$ radius of the cylinder. As shown in refs. [21,22], $B(r)$ is the solution of the differential equation

$$\nabla^2 B + K^2 B = 0$$

with the boundary condition $B(d) = H_a$ and $K$ given by

$$K^2 = \frac{2i}{\delta^2}\left(1 - \frac{n_s}{n_0}\right) - \frac{1}{\lambda_L^2} \qquad (2)$$

where $\delta = c/\sqrt{2\pi\sigma\omega}$ is the skin depth related to the electrical conductivity $\sigma$, $c$ the speed of light, $n_s$ the density of superconducting electrons, $n_0$ the total electron density and $\lambda_L = \sqrt{mc^2/4\pi n_s e^2}$ the London penetration depth. The real part of susceptibility $\chi$ is then given by [21]

$$\chi' = \mathrm{Re}\{4\pi^{-1}[-1 + 2J_1(Kd)/KdJ_0(Kd)]\} \qquad (3)$$

where $J_1$ and $J_0$ are Bessel functions. At low frequencies used in our experiments (all measurements presented in the main text were taken at $\omega/2\pi$=8 Hz) the skin depth is $\delta \gg d \gg \lambda_L$ ($\delta$ ~10 cm and $\lambda_L \approx 60$ nm), so that the first term in (2) can be neglected and $K^2$ becomes

$$K^2 \approx -\frac{1}{\lambda_L^2} = -\frac{4\pi n_s e^2}{mc^2}. \qquad (4)$$

Accordingly, $|Kd|$ in (3) can be replaced with $d/\lambda_L$, so that for a given $H_a$ the real part of susceptibility $\chi'$ is determined largely by the ratio of the superconductor's size and the magnetic field penetration depth $\lambda_L$. At and below $H_{c2}$, a continuous surface sheath screens the whole interior of the superconductor (as long as $\kappa$~1), resulting in perfect diamagnetic screening of small ac fields up to $H_{c2}$, i.e. $\chi' = -1/4\pi$ [23,24]. Such perfect screening, similar to the Meissner state, is seen for $H < H_{c2}$ in our measurements (Figs 1c and 2a in the main text). At $H_a > H_{c2}$, the order parameter $|\varphi|$ and therefore the superfluid fraction $\overline{n_s} = n_s/n_0$ become gradually suppressed [26], resulting in a gradual decrease of the diamagnetic susceptibility following eq. (2) (recall that $\overline{n_s} = |\varphi|^2$ (ref. [27]). The magnetic field dependence of susceptibility, $\chi'(H_a)$, is then determined by the evolution of the order parameter $|\varphi|^2(H_a)$, or $\overline{n_s}(H_a)$, with the magnetic field.

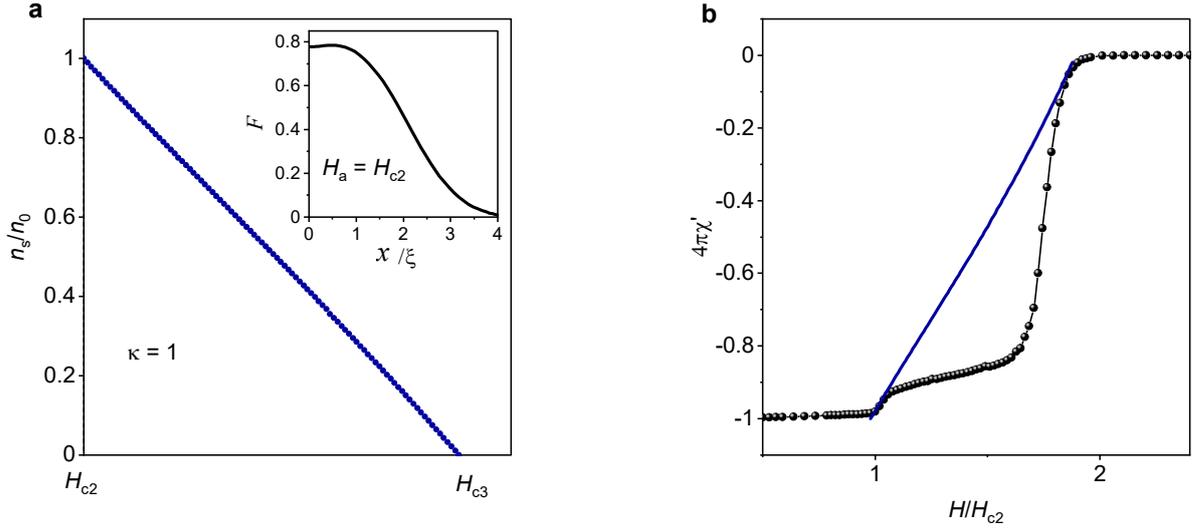

**Figure S10 | Magnetic field dependence of the calculated superfluid density and ac susceptibility for a conventional superconductor and comparison with experimental susceptibility for In$_2$Bi. a,** Superfluid fraction in the surface superconducting sheath, $n_s/n_0$, as a function of the external magnetic field $H_a$. *Inset*: spatial variation of the normalised order parameter $F(x)$ at $H_a = H_{c2}$ for GL parameter κ=1. The value of the order parameter at the surface, $F(0)$, is determined by requiring that $F(x)$ converges to zero inside the superconductor's bulk ($x \to \infty$). The calculated value of $F(0) \approx 0.8$ reproduces the result of ref. [26]. **b,** Comparison of the experimental ac susceptibility χ′ for an In$_2$Bi cylinder (black symbols, data of Fig. 1c in the main text) with χ′($H$) calculated from $n_s/n_0$ ($H_a$) in **a** (blue line).

To compare the expected dependence χ′($H_a$) with our data at $T = 2$K (where $n_s \approx n_0$ at and below $H_{c2}$), we followed ref. [26] where exact solutions of the Ginzburg-Landau equations were obtained for surface superconductivity as a function of $H_{c2} \leq H_a \leq H_{c3}$ and κ. First, we calculated numerically the normalised amplitude of the order parameter in the surface sheath, $F(H_a)$, where $F$ is defined as $F(qx) = D(qx)(|D_0|)^{-1}$, $D(qx)$ and $D_0$ stand for the amplitude of the order parameter at position $x$ in applied field $H_a$ and in zero field, respectively, and $qx = (x/\xi)(H_a/H_{c2})^{1/2}$, see ref. [26] for details. Using the corresponding equations in ref. [26], we found the spatial variation of $F(x/\xi)$ within the surface superconducting layer for different $H_a$, with the corresponding $\int_0^{4\xi} F^2(t)dt$ proportional to the superfluid fraction $\overline{n_s}$ in the surface sheath. The inset of Fig. S10a shows the result at $H_a = H_{c2}$ (reproducing the calculations in ref. [26]) and the main panel shows the corresponding $H$ dependence of $\overline{n_s}$.

The obtained results for $n_s/n_0(H_a)$ were then used to calculate the susceptibility χ′($H_a$) using (3) and (4). The result is shown in Fig. S10b by the blue line: χ′ is expected to decrease approximately linearly as the field increases from $H_{c2}$ to $H_{c3}$. Such a smooth, near-linear dependence of χ′ is in agreement with observations on high-quality Nb in literature (e.g., ref. [28]). A similar smooth χ′($H_a$) dependence was also observed in our In$_2$Bi samples with a roughened surface and at relatively large amplitudes of the ac field (Figs 3b and 2a, respectively). It is however in stark contrast to our observations for In$_2$Bi crystals with smooth surfaces at small ac amplitudes, where χ′ changes little up to $H_{ts}$ (Fig. S10b).

DC magnetization measurements probe the total magnetic moment in the sample, therefore the contribution from the surface sheath can only be seen at and above $H_{c2}$ (below $H_{c2}$ it is masked by bulk magnetization). As discussed in detail in refs [23,25], for a cylinder of radius $R$, magnetization per unit volume $M$ and the maximum current that the sheath can sustain are size dependent and inversely proportional to $R^{1/2}$, i.e., the larger the radius, the smaller the current and the magnetic moment, due to the 'cost' in magnetization energy for the whole volume of the cylinder. Nevertheless, $M$ is always much larger [by a factor proportional to $(R/\lambda)^{1/2}$] than the equilibrium moment $M_0$ corresponding to zero total current in the sheath. As the field is increased above $H_{c2}$, $M$ decreases due to a decrease in the free energy of the sheath [25].

## 9. Effect of topological surface states on surface superconductivity

To find how the presence of the topological surface states modifies the properties of the superconducting surface sheath we extend the existing theory of surface superconductivity [26]. The latter yields the spatial distribution of the order parameter $\varphi(\boldsymbol{r})$ and its evolution with the external dc field as it varies between $H_{c2}$ and $H_{c3}$. As shown in ref. [26] (and reproduced in the inset of Fig. S10a) the order parameter in conventional superconductors remains constant over a distance $x \approx \xi$ from the surface and decays exponentially to zero at $x \approx 4\xi$. At $H_{c2}$ the order parameter at the surface is close to its bulk value but, as the external field increases, it decreases approximately linearly and vanishes at $H_{c3}$. This result [26] was obtained by solving the Ginzburg-Landau equations with the conventional boundary condition for a superconductor-dielectric interface, that is, that the derivative of the order parameter vanishes at the superconductor's surface [29].

Of particular relevance to our experimental observations is the effect of the superconducting topological surface states on the superfluid density (fraction of the superconducting electrons) $\overline{n_s} = |\varphi|^2$ as a function of the applied dc magnetic field. As shown below, their presence changes the boundary condition, greatly enhancing the overall $\overline{n_s}$ in the surface sheath as compared to the case of conventional surface superconductivity and concentrating the screening currents in a narrow region at the surface. This results in a much more robust screening of magnetic field above $H_{c2}$.

Following ref. [26], we consider the Ginzburg-Landau free energy

$$\mathcal{F}[\varphi, \boldsymbol{A}] = \int d\boldsymbol{r} \left\{ -\alpha \, |\varphi(\boldsymbol{r})|^2 + \frac{\beta}{2} \, |\varphi(\boldsymbol{r})|^4 + \frac{[\boldsymbol{\nabla} \times \boldsymbol{A}(\boldsymbol{r})]^2}{8\pi} + \frac{1}{2m^*} \left| -i\hbar \boldsymbol{\nabla} \varphi(\boldsymbol{r}) + \frac{e^*}{c} \boldsymbol{A}(\boldsymbol{r})\varphi(\boldsymbol{r}) \right|^2 \right\},$$

where $\alpha = \hbar^2/(2m^*\xi^2)$, $\xi$ is the coherence length, $\beta = 4\pi\alpha^2/H_c^2$, $H_c$ the thermodynamic critical field, $\varphi(\boldsymbol{r})$ the superconducting order parameter, $\boldsymbol{A}(\boldsymbol{r})$ the vector potential, $m^*$ the particle mass and $e^* = 2e$. Here $e$ is the electron charge. In the Meissner state, the equilibrium order parameter is $\bar{\varphi} = -\alpha/b$. The London penetration length is $\lambda = 4\pi(e^*|\bar{\varphi}|)^2/(m^*c^2)$, and the Ginzburg-Landau parameter $\kappa = \lambda/\xi$.

We consider a planar geometry such that the superconductor occupies the half space $x > 0$, while the topological surface state is assumed to have zero thickness and located at $x = 0$. To derive equations amenable for numerical solution, we rescale the lengths with $\xi\mu$, where $\mu = \sqrt{H_{c2}/H_{dc}}$, and introduce a dimensionless order parameter $F(\zeta)$ such that $\varphi(\zeta) = \bar{\varphi} \, F(\zeta)e^{ik\varsigma}$. $F(\zeta)$ is a real function, whereas $\zeta = x/\xi\mu$ and $\varsigma = y/\xi\mu$ are dimensionless variables. Similarly, the vector potential is rescaled with $\sqrt{2}\lambda H_c/\mu$. In these equations, $H_{dc}$ is the applied dc magnetic field and $H_{c2} = \sqrt{2}\kappa H_c$.

We further introduce [26] a constant $a_0$ and the function $a(\zeta)$ to rewrite the dimensionless vector potential as $\zeta + a_0 + a(\zeta)$. Here, the first term is the vector potential due to the applied dc field, and $a_0$ is the total vector potential at the superconductor's surface. Accordingly, $a(0) = 0$. In terms of these functions, the free energy becomes

$$\mathcal{F}[F,a] = \frac{\xi S H_c^2}{4\pi\mu^2} \int_0^\infty d\zeta \left\{ \mu^2 \left( \frac{F^2(\zeta)}{2} - 1 \right) F^2(\zeta) + \frac{\kappa^2}{\mu^2}[1 + \partial_\zeta a(\zeta)]^2 + [\partial_\zeta F(\zeta)]^2 \right.$$
$$\left. + [\zeta - \Gamma + a(\zeta)]^2 F^2(\zeta) \right\},$$

where $\Gamma = k - a_0$ (recall that $k$ is the wavevector that controls the phase of the order parameter) and $S$ is the area of the superconductor's surface. Taking the functional derivative of $\mathcal{F}[F,a]$ with respect to $F(\zeta)$ and $a(\zeta)$ we find the Ginzburg-Landau equations

$$\begin{cases} \partial_\zeta^2 F(\zeta) + \mu^2 F^2(\zeta)[1 - F^2(\zeta)] - [\zeta - \Gamma + a(\zeta)]^2 F(\zeta) = 0 \\ \quad \frac{\kappa^2}{\mu^2} \partial_\zeta^2 F(\zeta) = F^2(\zeta)[\zeta - \Gamma + a(\zeta)] \end{cases} \qquad (5).$$

These equations must be complemented by the following constraint

$$\Gamma = \mu \sqrt{1 - \frac{F^2(0)}{2} + [\partial_\zeta F(\zeta)]^2_{\zeta=0}} \qquad (6),$$

which is obtained by minimizing the free energy with respect to $\Gamma$ [26]. By imposing the appropriate boundary conditions at $\zeta = 0$ (see below), Eqs. (5) − (6) are solved to yield $F(\zeta)$ and $a(\zeta)$. In both ref.[26] and the present calculations, $a(0)$ and $\partial_\zeta a(x)|_{\zeta=0}$ are required to vanish at $\zeta = 0$, while $F(\zeta)$ and $a(\zeta)$ must converge to a constant deep in the superconductor's bulk ($\zeta \to \infty$). In particular, $F(\zeta \to \infty) = 0$ (recall that we consider the case of $H_{dc} > H_{c2}$). In ref. [26], the set of Eqs. (5) − (6) was solved by imposing the conventional (Abrikosov) boundary condition [29] for a superconductor-dielectric interface, which is $\partial_\zeta F(\zeta)|_{\zeta=0} = 0$. $F(0)$ was then determined with a shooting method so that the function $F(\zeta)$ satisfied all the above requirements.

To reflect the presence of the topological surface states in our experiment, we employ a different boundary condition. Physically, we must require that the total current across the boundary vanishes. This is generally satisfied by imposing [27]

$$\partial_\zeta F(\zeta)|_{\zeta=0} = \frac{F(0)}{b},$$

where $b$ is a constant. For a superconductor-dielectric interface, $b \to \infty$. For a boundary between two superconductors, $b$ can be finite and negative. This should happen if one of the superconductors is a film with thickness much smaller than the magnetic field penetration depth $\lambda$ and, therefore, its order parameter is little affected by the applied magnetic field [30].

Our experimental system can be thought of as an atomically-thin metallic film of thickness $d_{ts}$, which surrounds a superconducting cylinder. Accordingly, we model the system as two mutually proximitized superconductors, one corresponding to bulk $In_2Bi$ and the other to its topological surface state. In a magnetic field parallel to the surface, the critical field of the film itself must be strongly enhanced because of its atomic-scale thickness [27] (recall that the thickness of the surface sheath due to the superconducting bulk is relatively large: $d \approx 2\xi > \lambda$, $\xi \approx 60$ nm and $\lambda \approx 65$ nm; see main text). Being so thin, the topological superconducting film is expected to be essentially unaffected by $H_{dc} < H_{c3}$, which allows us to assume $F(0) = 1$ for all relevant values of $H_{dc}$. In turn, this 'pinned' order parameter is found to modify the boundary condition for the surface superconducting sheath associated with the bulk superconductor.

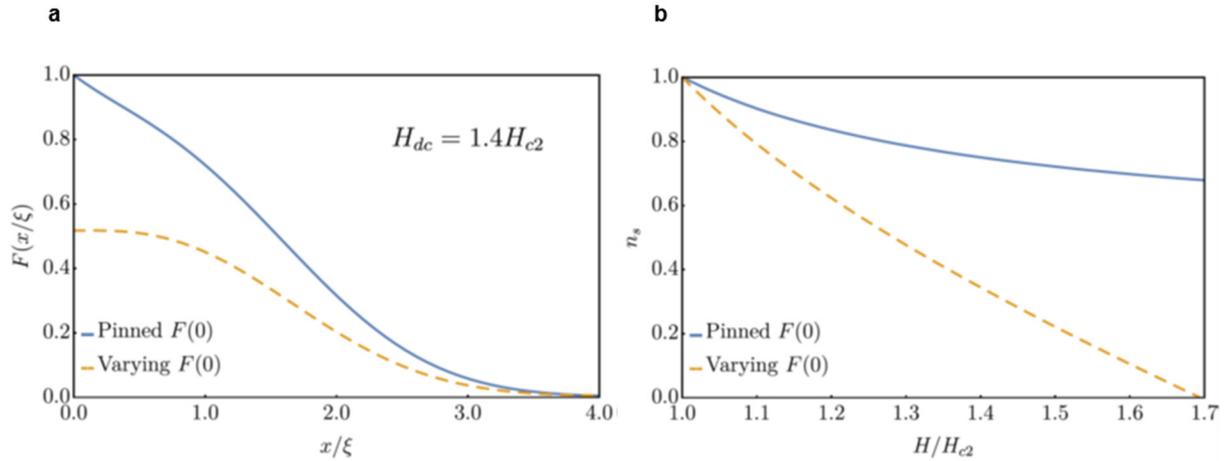

**Figure S11 | Solutions of the Ginzburg-Landau equations with and without the superconducting topological state at the surface. a,** Comparison between the conventional behaviour [varying $F(0)$] and our case of $F$ 'pinned' to unity at the surface. **b,** The corresponding superfluid densities. In the 'pinned' case, the surface superconductivity is less effected by applied field $H$.

Based on these considerations, we solve the Ginzburg-Landau equations $(5) - (6)$ by imposing $F(0) = 1$ and determine the resulting $b$ by a shooting method so that $F(\zeta \to \infty) = 0$ and $a(\zeta \to \infty)$ converges to a constant. The resulting order parameter is shown in Fig. S11a. One can see that, for a given field $H_{c2} < H < H_{c3}$, $F$ is considerably enhanced in the whole $\sim 4\xi$ surface layer, compared to the conventional behaviour for the surface superconductivity [26], which is also shown in Fig. S11a. As a consequence, the superfluid density above $H_{c2}$, $n_s \propto \int_0^\infty d\zeta\, F^2(\zeta)$, decreases much slower with increasing the magnetic field than in the conventional case [26] (Fig. S11b). Because above $H_{c2}$ the experimentally measured ac susceptibility is determined by $\overline{n_s}(H/H_{c2})$ (see the previous section), the results of Fig. S11b can be translated directly into the susceptibility. The resulting behaviour is plotted in Fig. 4d (main text) showing good agreement between the experiment and theory.

Note that the Ginzburg-Landau free energy is generally reduced if $F(\zeta)$ has a negative slope at $x = 0$. Therefore, it is plausible to argue that the system will always try to minimize its energy by realising an order parameter that peaks at $x = 0$. The presence of the topological surface state opens up such a possibility, and the system readily adapts. For analysis beyond the phenomenological Ginzburg-Landau equations, it would require considering the microscopic interplay between the order parameters in the surface sheath and topological states, which involves self-consistent solution of the Gor'kov equation [29]. This feat is beyond the scope of the present work.

## References for Supporting Information